\documentclass[11pt]{article}
\usepackage{amssymb}
\usepackage{epsfig}
\usepackage{graphicx}

\newcommand{\BE}{\begin{equation}}
\newcommand{\EE}{\end{equation}}

\usepackage{amsmath}
\begin{document}
\begin{titlepage}

\vspace*{1mm}
\begin{center}

\vskip 1 pt

  {\LARGE{\bf {Cosmic Microwave Background and the issue of a fundamental preferred  frame}}}

\end{center}

\begin{center}

\vspace*{14mm} {\Large  M. Consoli$^{(a)}$ and A. Pluchino
$^{(b,a)}$}
\vspace*{4mm}\\
{a) Istituto Nazionale di Fisica Nucleare, Sezione di Catania, Italy ~~~~~~~~~\\
b) Dipartimento di Fisica e Astronomia dell'Universit\`a di Catania,
Italy }
\end{center}

\begin{center}
{\bf Abstract}
\end{center}


\par\noindent The possibility to correlate ether-drift measurements
in laboratory and direct CMB observations with satellites in space
would definitely confirm the existence of a fundamental preferred
frame for relativity. Today, the small residuals observed so far
(from Michelson-Morley onward) are just considered typical
instrumental effects in experiments with better and better
sensitivity. Though, if the velocity of light propagating in the
various interferometers is not exactly the same parameter $c$ of
Lorentz transformations, nothing would really prevent to observe an
ether drift. Thus, for the earth cosmic velocity v=370 km/s, we
argue that a fundamental $10^{-15}$ light anisotropy, as presently
observed in vacuum and in solid dielectrics, is revealing a
$10^{-9}$ difference in the effective vacuum refractivity between an
apparatus in an ideal freely-falling frame and an apparatus on the
earth surface. In this perspective, the stochastic nature of the
physical vacuum could also explain the irregular character of the
signal and the observed substantial reduction from its instantaneous
$10^{-15}$ value to its statistical average $10^{-18}$ (or smaller).
For the same v=370 km/s the different refractivities, respectively
${\cal O}(10^{-4})$ and ${\cal O}(10^{-5})$ for air or helium at
atmospheric pressure, could also explain the observed light
anisotropy, respectively ${\cal O}(10^{-10})$ and ${\cal
O}(10^{-11})$. However, for consistency, one should also understand
the physical mechanism which enhances the signal in weakly bound
gaseous matter but remains ineffective in solid dielectrics where
the refractivity is ${\cal O}(1)$. This mechanism is naturally
identified in a non-local, tiny temperature gradient of a fraction
of millikelvin which is found in all classical experiments and might
ultimately be related to the CMB temperature dipole of $\pm 3$ mK or
reflect the fundamental energy flow associated with a
Lorentz-non-invariant vacuum state. The importance of the issue
would deserve more stringent tests  with dedicated experiments and
significant improvements in the data analysis.

 \vskip 15 pt



\end{titlepage}


\section{Introduction}

\subsection{CMB and the issue of a fundamental preferred frame}

Precise observations with satellites in space have revealed a tiny
anisotropy in the temperature of the Cosmic Microwave Background
(CMB) \cite{mather,smoot}. The present interpretation of its
dominant dipole component (the {\it kinematic} dipole \cite{yoon})
is in terms of a Doppler effect ($\beta=v/c)$\BE
T(\theta)={{T_o\sqrt{1-\beta^2}}\over{1- \beta \cos \theta} } \EE
due to a motion of the solar system with average velocity $v\sim
370$ km/s toward a point in the sky of right ascension $\alpha \sim
168^o$ and declination $\gamma\sim -7^o$. Therefore, if one sets
$T_o \sim $ 2.7 K and $\beta\sim 0.0012$, as for $v\sim$ 370 km/s,
there are angular variations of a few millikelvin \BE
\label{CBR}\Delta T(\theta) \sim T_o \beta \cos\theta \sim \pm 3
~{\rm mK} \EE

By accepting this interpretation, the question naturally arises
concerning the role of the system where the CMB dipole anisotropy
vanishes exactly, i.e. does it represent a fundamental frame for
relativity? The usual answer is that the two concepts are unrelated.
Namely, the CMB is a definite medium with a rest frame where the
dipole anisotropy is zero. Motion with respect to this frame can be
detected and, in fact, has been detected. But the existence of a
fundamental preferred frame would contradict special relativity
which is the presently accepted interpretation of the theory.

Still, it should not be overlooked that the observed CMB dipole can
be reconstructed, to good approximation, by combining the various
peculiar motions which are involved, namely the rotation of the
solar system around the galactic center, the motion of the Milky Way
around the center of the Local Group and the motion of the Local
Group of galaxies in the direction of that large concentration of
matter known as the Great Attractor \cite{smoot}. In this way, once
a vanishing CMB dipole is equivalent to switching-off all possible
peculiar motions, one naturally arrives to the concept of a global
frame of rest determined by the average distribution of matter in
the universe.

At the same time, at a more formal level, the idea of a preferred
frame finds other motivations in the modern picture of the vacuum,
intended as the lowest energy state of the theory. This is believed
to arise from the macroscopic condensation process, see e.g.
\cite{thooft,mech,foop}, of elementary quanta (Higgs particles,
quark-antiquark pairs, gluons...) in the same zero-3-momentum state
and thus, by definition, singles out some reference frame $\Sigma$.
Then, the fundamental question \cite{foop} is how to reconcile this
picture with a basic postulate of axiomatic quantum field theory:
the exact Lorentz invariance of the vacuum \cite{cpt}.

Usually this is not considered as a problem with the motivation,
perhaps, that the average properties of the condensed phase are
summarized into a single quantity which transforms as a world scalar
under the Lorentz group, for instance, in the Standard Model, the
vacuum expectation value $\langle\Phi\rangle$ of the Higgs field.
However, this does not necessarily imply that the vacuum state
itself has to be {\it Lorentz invariant}. Namely, Lorentz
transformation operators ${U}'$, ${U}''$,..could transform non
trivially the reference vacuum state \footnote{ We ignore here the
problem of vacuum degeneracy by assuming that any overlapping among
equivalent vacua vanishes in the infinite-volume limit of quantum
field theory (see e.g. S. Weinberg, {\it The Quantum Theory of
Fields}, Cambridge University press, Vol.II, pp. 163-167).}
$|\Psi^{(0)}\rangle$ (appropriate to an observer at rest in
$\Sigma$) into $| \Psi'\rangle$, $| \Psi''\rangle$,.. (appropriate
to moving observers $S'$, $S''$,..) and still, for any
Lorentz-invariant operator ${G}$, one would find \BE \langle
{G}\rangle_{\Psi^{(0)}}=\langle {G}\rangle_{\Psi'}=\langle
{G}\rangle_{\Psi''}=..\end{equation} For the convenience of the
reader, we will report in the Appendix the main ingredients to
understand the origin of the problem (see also \cite{foop}).  Here,
in this Introduction, we will only limit ourselves to the general
conclusion: imposing that only scalar fields can acquire a
non-vanishing vacuum expectation value does {\it not} guarantee that
the vacuum state itself is Lorentz invariant.

This preliminary discussion is important because, on this basis, the
mentioned global frame could also reflect a vacuum structure with
some degree of substantiality and thus characterize non-trivially
the form of relativity which is physically realized in nature. In
other words, as in the original Lorentzian formulation, Lorentz
transformations could still be exact to connect two observers in
uniform translational motion but there would be a preferred
reference frame. In this case, the isotropy of the CMB radiation
would just {\it indicate} the existence of such a global frame that
we could decide to call the ``ether'', but the cosmic radiation
itself would {\it not} coincide with this type of ether. Ultimate
implications are far reaching. Think for instance of the
possibility, with a preferred frame, to reconcile faster-than-light
signals with causality \cite{annals} and thus provide a very
different view of the apparent non-local aspects of the quantum
theory \cite{scarani} \footnote{The importance of establishing a
link between CMB and ether-drift experiments is better illustrated
by quoting from ref.\cite{hardy} where Hardy discusses the
implications of the typical non-locality of the quantum theory:
``Thus, Nonlocality is most naturally incorporated into a theory in
which there is a special frame of reference. One possible candidate
for this special frame of reference is the one in which the cosmic
background radiation is isotropic. However, other than the fact that
a realistic interpretation of quantum mechanics requires a preferred
frame and the cosmic background radiation provides us with one,
there is no readily apparent reason why the two should be
linked''.}.

Therefore, once the answer to our basic question cannot be found
with theoretical arguments only, one should look at experiments, in
particular at the so called ``ether-drift'' experiments where one
tries to measure a small difference of the velocity of light in
different directions and, eventually, to find a definite correlation
with the direct CMB observations in space. At present, the general
consensus is that no physical ether drift has ever been detected.
This standard view considers all available data (from
Michelson-Morley until the modern interference experiments with
optical resonators) as a long sequence of null results, i.e. typical
instrumental effects obtained in measurements with better and better
systematics.

However, by accepting the idea of a preferred reference frame, and
if the velocity of light propagating in the various interferometers
is not {\it exactly} the same parameter $c$ of Lorentz
transformations, nothing would really prevent to observe an ether
drift. For this reason, it is natural to enquire to which extent the
experiments performed so far have really given null results. By
changing the theoretical model, small residual effects which
apparently represent spurious instrumental artifacts could acquire a
definite physical meaning with substantial implications for both
physics and the history of science.

\subsection{Basics of the ether-drift experiments}

To introduce the argument, it is necessary to recall the basic
ingredients of the ether-drift experiments starting from the
original 1887 Michelson-Morley experiment \cite{mm}. Nowadays, this
fundamental experiment and its early repetitions performed  at the
turn of 19th and 20th centuries (by Miller \cite{miller}, Kennedy
\cite{kenconference}, Illingworth \cite{illingworth}, Joos
\cite{joos}...) are considered as a venerable, well understood
historical chapter for which, at least from a physical point of
view, there is nothing more to refine or clarify. All emphasis is
now on the modern versions of these experiments, with lasers
stabilized by optical cavities, see e.g. \cite{applied} for a
review. These modern experiments adopt a very different technology
but, in the end, have exactly the same scope: searching for the
possible existence of a preferred reference frame through an
anisotropy of the two-way velocity of light
$\bar{c}_\gamma(\theta)$. This is the only one that can be measured
unambiguously and is defined in terms of the one-way velocity
$c_\gamma(\theta)$ as
\begin{equation}
\label{first0}
\bar{c}_\gamma(\theta)={{2~c_\gamma(\theta)c_\gamma(\pi
+\theta)}\over{c_\gamma(\theta) +c_\gamma (\pi +\theta)}}
\end{equation}
Here $\theta$ represents the angle between the direction of light
propagation and the earth velocity with respect to the hypothetical
preferred frame $\Sigma$. By introducing the anisotropy
\[ \Delta\bar{c}_\theta =
\bar{c}_\gamma(\pi/2+\theta)-\bar{c}_\gamma(\theta) \] there is a
simple relation with the time difference $\Delta t(\theta)$ for
light propagation back and forth along perpendicular rods of length
$D$. In fact, by assuming the validity of Lorentz transformations,
the length of a rod does not depend on its orientation, in the $S'$
frame where it is at rest, and one finds

\begin{figure}
\includegraphics[bb=-300 0 0 450,
angle=0,scale=0.35] {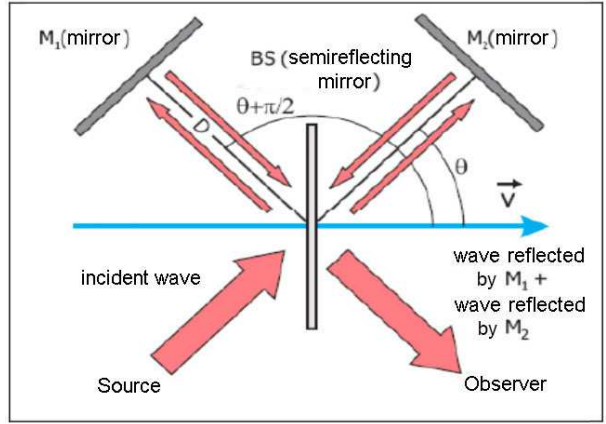}
\caption{\it The typical scheme of Michelson's interferometer.}
\label{Michinterferometer}
\end{figure}

\begin{equation}
\label{deltaT} \Delta t(\theta)=
{{2D}\over{\bar{c}_\gamma(\theta)}}-
{{2D}\over{\bar{c}_\gamma(\pi/2+\theta)}} \sim
{{2D}\over{c}}~{{\Delta \bar{c}_\theta } \over{c}}
\end{equation}
(where, in the last relation, we have assumed that light propagates
in a medium of refractive index ${\cal N}=1 + \epsilon$, with
$\epsilon\ll 1$). This gives directly the fringe patterns ($\lambda$
is the light wavelength)
\begin{equation}
\label{newintro} {{\Delta \lambda(\theta)}\over{\lambda}} \sim
{{2D}\over{\lambda}} ~{{\Delta \bar{c}_\theta } \over{c}}
\end{equation}
which were measured with Michelson interferometers in classical
ether-drift experiments, see Fig.\ref{Michinterferometer}.

In modern experiments, on the other hand, a possible anisotropy of
$\bar{c}_\gamma(\theta)$ would show up through the relative
frequency shift, i.e. the beat signal, $\Delta\nu(\theta)$ of two
orthogonal optical resonators. Their frequency
\begin{equation}
\label{nutheta0}
  \nu (\theta)= {{ \bar{c}_\gamma(\theta) m}\over{2L}}
\end{equation}
is proportional to the two-way velocity of light within the
resonator through an integer number $m$, which fixes the cavity
mode, and the length of the cavity $L$ as measured in the laboratory
$S'$ frame. Therefore, once the length of a cavity in its rest frame
does not depend on its orientation, one finds
\begin{equation}
\label{bbasic2}
 {{\Delta \nu(\theta) }\over{\nu_0}}  \sim
      {{\Delta \bar{c}_\theta } \over{c}}
\end{equation}
where $\nu_0$ is the reference frequency of the two resonators.

Within this basic scheme, let us now see how experimental results
are presented, in particular in the recent work of Nagel et al.
\cite{nagelnature}. Their measurements amount to an {\it average}
fractional anisotropy $|\langle
{{\Delta\bar{c}_\theta}\over{c}}\rangle| \lesssim 10^{-18}$. With
this new result, by looking at their Fig.1 where all ether-drift
experiments are reported, one gets the impression of a steady,
substantial improvement over the original 1887 Michelson-Morley
result ${{|\Delta\bar{c}_\theta|}\over{c}}={\cal O}(10^{-10})$. All
together, their plot supports the mentioned view of a series of null
results with better and better systematics.

Still, over the years, greatest experts \cite{hicks,miller} have
seriously questioned the traditional null interpretation of the very
early measurements. In their opinion, the small residuals should not
be neglected. Therefore one may wonder if, indeed, this first
impression is correct. For instance, the various measurements were
performed in different conditions, i.e. with light propagating in
gaseous media (as in
\cite{mm,miller,kenconference,illingworth,joos}) or in a high vacuum
(as in \cite{brillet,newberlin,newschiller,cpt2013,schiller2015}) or
inside dielectrics with a large refractive index (as in
\cite{fox,nagelnature}) and there could be physical reasons which
prevent a straightforward comparison. In this case, the difference
between old experiments (in air or gaseous helium) and modern
experiments (in vacuum or solid dielectrics) might not depend on the
technological progress only but also on the different media that
were tested.

With this perspective, let us re-consider Maxwell's classical
calculation \cite{maxwell} of light anisotropy with a preferred
frame which was at the base of the ether-drift experiments.  We know
that his original estimate, namely
${{|\Delta\bar{c}_\theta|}\over{c}}\sim \beta^2$, is wrong. However,
even by assuming the exact validity of Lorentz transformations,
Maxwell's problem still makes sense. The point is that, for a
refractive index ${\cal N} =1+\epsilon $, in the $\epsilon \to 0$
limit where the velocity of light tends to coincide with the basic
parameter $c$ entering Lorentz transformations, simple symmetry
arguments suggest that a possible non-zero result has the form
${{|\Delta\bar{c}_\theta|}\over{c}}\sim \epsilon \beta^2$. From this
relation and by assuming the typical value $v\sim 370$ km/s for the
earth cosmic motion, for air at room temperature and atmospheric
pressure where the refractivity is ${\cal N}- 1 \sim 2.8 \cdot
10^{-4} $ and for gaseous helium at room temperature and atmospheric
pressure where the refractivity is ${\cal N}-1\sim 3.3 \cdot
10^{-5}$, one gets anisotropy values, respectively
${{|\Delta\bar{c}_\theta|}\over{c}}= {\cal O}(10^{-10})$ and
${{|\Delta\bar{c}_\theta|}\over{c}}= {\cal O}(10^{-11})$, which are
much smaller than the classical prediction
${{|\Delta\bar{c}_\theta|}\over{c}} \sim 10^{-8}$ (for the
traditional orbital value $v=$ 30 km/s) and consistent with the
actual observations.

At the same time, symmetry arguments give often a successful
description of phenomena independently of the particular physical
mechanisms. As such, this view does not necessarily contradict the
standard interpretation of the residuals as thermal disturbances.
Indeed, as pointed out by Kennedy \cite{kenconference}, also these
disturbances become smaller and smaller for $\epsilon \to 0$. For
this reason, periodic temperature variations of $1\div 2$
millikelvin in the air of the optical arms, were considered by
Shankland et al. \cite{shankland}, Kennedy (see p. 175 of
ref.\cite{shankland}) and Joos \cite{joos2} to explain away Miller's
observations, but were never fully understood. As such, these
thermal effects might have a {\it non-local} origin somehow
associated with an earth velocity $v$, e.g. our motion within the
CMB. Finding such an explanation, where symmetry arguments, on the
one hand, motivate and, on the other hand, find justification in
underlying physical mechanisms, would greatly increase our
understanding. We will return to this crucial point in the
following.

\subsection{Nature of the vacuum and time dependence of the data}

After the magnitude of the signal, the other important aspect of the
ether-drift experiments concerns the time dependence of the data.
Traditionally, it has always been assumed that, for short-time
observations of a few days, where there are no sizeable changes in
the earth orbital velocity, the time dependence of a genuine
physical signal should reproduce the slow and regular modulations
induced by the earth rotation. The data instead, for both classical
and modern experiments, have always shown a very irregular behavior.
As a consequence, all statistical averages are much smaller than the
instantaneous values. For instance, compare the average anisotropy
$|\langle {{\Delta\bar{c}_\theta}\over{c}}\rangle| \lesssim
10^{-18}$ obtained in ref.\cite{nagelnature} by combining a large
number of observations with its typical, {\it instantaneous}
determination ${{|\Delta\bar{c}_\theta|}\over{c}}\lesssim 10^{-15}$
shown in their Fig.3 b. This difference, between single measurements
and statistical averages, has always represented a strong argument
to interpret the data as mere instrumental artifacts.

However, again, could there be an alternative interpretative scheme?
In other words, could a definite {\it instantaneous} value $
{{\Delta\bar{c}_\theta}\over{c}} \neq 0$ coexist with its vanishing
statistical average $|\langle
{{\Delta\bar{c}_\theta}\over{c}}\rangle|$? This possibility was
considered in refs.\cite{plus,physica} by modeling the physical
vacuum as a fundamental stochastic medium, somehow similar to an
underlying turbulent fluid.

To understand the motivations, let us observe that it is the nature
of the physical vacuum to determine the relation between the
macroscopic earth motion and the microscopic optical measurements in
a laboratory. Light propagation (e.g. inside an optical cavity)
takes place in this substratum which is dragged along the earth
motion but, so to speak, is not rigidly connected with the solid
parts of the apparatus as fixed in the laboratory. Therefore, if one
would try to characterize its local state of motion, say $v_\mu(t)$,
this does not necessarily coincide with the projection of the global
earth motion, say $\tilde v_\mu (t)$, at the observation site. The
latter is a smooth function while the former, $v_\mu(t)$, in
principle is unknown. By comparing with the motion of a body in a
fluid, the equality $v_\mu(t)=\tilde v_\mu(t)$ amounts to assume a
form of regular, laminar flow where global and local velocity fields
coincide. Instead, in a turbulent fluid large-scale and small-scale
flows would only be {\it indirectly} related.

The simplest explanation for the turbulent-fluid analogy is the
intuitive representation of the vacuum as a fluid with vanishing
viscosity. Then, in the framework of the Navier-Stokes equation, a
laminar flow is by no means obvious due to the subtlety of the
zero-viscosity (or infinite Reynolds number) limit, see for instance
the discussion given by Feynman in Sect. 41.5, Vol.II of his
Lectures \cite{feybook}. The reason is that the velocity field of
such a hypothetical fluid cannot be a differentiable function
\cite{onsager} and one should think, instead, in terms of
continuous, nowhere differentiable functions, similar to ideal
Brownian paths \cite{eyink}. This gives the idea of the vacuum as a
fundamental stochastic medium consistently with some basic
foundational aspects of {\it both} quantum physics and relativity
\footnote{This picture was first proposed in the old ether theory at
the end of XIX Century \cite{whittaker}. In this original
derivation, the Lorentz covariance of Maxwell equations was not
postulated from scratch but was emerging from an underlying physical
system whose constituents obey classical mechanics. More recently,
the turbulent-ether model has been re-formulated by Troshkin
\cite{troshkin} (see also \cite{puthoff} and \cite{tsankov}) in the
framework of the Navier-Stokes equation and by Saul \cite{saul} by
starting from Boltzmann's transport equation. As another example,
the same picture of the physical vacuum (or ether) as a turbulent
fluid was Nelson's \cite{nelson} starting point. In particular, the
zero-viscosity limit gave him the motivation to expect that ``the
Brownian motion in the ether will not be smooth'' and, therefore, to
conceive the particular form of kinematics which is at the base of
his stochastic derivation of the Schr\"odinger equation. A
qualitatively similar picture is also obtained by representing
relativistic particle propagation from the superposition, at very
short time scales, of non-relativistic particle paths with different
Newtonian mass \cite{kleinert} . In this formulation, particles
randomly propagate (in the sense of Brownian motion) in an
underlying granular medium which replaces the trivial empty vacuum
\cite{jizba} . For more details, see \cite{physica} .}.

On this basis, in refs.\cite{plus,physica} the signal was
characterized as in the simulations of turbulent flows. Namely, the
local $v_\mu(t)$ exhibits random fluctuations while the global
$\tilde v_\mu(t)$ determines its typical boundaries. Then, if
turbulence becomes homogeneous and isotropic at small scales, as it
is generally accepted in the limit of zero viscosity, the direction
of the local drift becomes a completely random quantity which has no
definite limit by combining a large number of observations. Since
vectorial quantities have vanishing statistical averages, one should
thus analyze the data in phase and amplitude (which give
respectively the instantaneous direction and magnitude of the local
drift) and concentrate on the latter which is a positive-definite
quantity and remains non-zero under any averaging procedure. In this
alternative picture of the ether drift, a non-vanishing amplitude
(i.e. definitely larger than the experimental resolution) becomes
the signature to separate an irregular, but genuine, physical signal
from spurious instrumental noise. If non-zero, its time modulations
can then be compared with models of the earth cosmic motion.

\subsection{Comparing experiments in gases, vacuum and solid dielectrics}

After these preliminaries, a definite quantitative framework to
analyze the experiments will be presented in Sects.2 and 3. Our
scheme can be considered a modern version of the original Maxwell
calculation. It applies to the infinitesimal region of refractive
index ${\cal N}= 1 + \epsilon$ and, within the analogy of a
turbulent flow, takes into account random fluctuations of the local
drift around the average earth motion. As a matter of fact, see
Sect.4, when this scheme is used to describe the small residuals of
the classical ether-drift experiments in gaseous systems, it yields
typical earth velocities which are consistent with the value of 370
km/s obtained from the CMB observations. Confirming (or excluding)
this alternative interpretation will thus require a new series of
dedicated experiments. The essential aspect is that the optical
resonators, which are coupled to the lasers, should be filled by
gaseous media. In this way, one could reproduce the experimental
conditions of those early measurements with today's much greater
accuracy. Such experiments would be along the lines of
ref.\cite{holger} where just the use of optical cavities filled with
different forms of matter was considered as a useful complementary
tool to study deviations from exact Lorentz invariance.

Waiting for this new series of experimental tests, however, we have
tried to check the same scheme in modern vacuum experiments. The
point is that for the {\it physical} vacuum the ideal equality
${\cal N}_v =1$ might not be exact. For instance, it was proposed in
\cite{gerg} that, if the curvature observed in a gravitational field
is an emergent phenomenon from a fundamentally flat space-time,
there should be a small vacuum refractivity $\epsilon_v={\cal N}_v -
1 \sim 10^{-9}$. This would take into account the difference which
exists, in this case, between an apparatus in an ideal freely
falling frame and an apparatus placed on the earth surface. The
basic argument will be repeated in Sect.5 with many additional
details not reported in \cite{gerg}. The existence of a preferred
frame would then imply in our picture a definite, instantaneous
${{|\Delta\bar{c}_\theta|}\over{c}} \sim \epsilon_v \beta^2\sim
10^{-15}$ which coexists with much smaller statistical averages
$|\langle {{\Delta\bar{c}_\theta}\over{c}}\rangle| \ll 10^{-15}$. In
Sect. 6, this expectation will be shown to be consistent with the
most recent room temperature and cryogenic vacuum experiments.

Then, in Sect.7, we will re-consider from scratch the temperature
dependence of the refractivity in gaseous media. This calculation
shows that the light anisotropy ${{|\Delta\bar{c}_\theta|}\over{c}}
\sim \epsilon\beta^2$ observed in gases could also be interpreted as
a thermal effect due to a tiny, non-local temperature gradient of a
fraction of millikelvin. This is found in all classical experiments
and might be related to the CMB temperature dipole of $\pm 3$ mK or
also reflect the fundamental energy flow expected in a
Lorentz-non-invariant vacuum. Whatever its ultimate origin, the
interesting point is that this thermal interpretation provides a
dynamical basis for the {\it enhancement} found in weakly bound
gaseous matter (i.e. the observed magnitudes
${{|\Delta\bar{c}_\theta|}\over{c}}= {\cal O}(10^{-10})$ and
${{|\Delta\bar{c}_\theta|}\over{c}}={\cal O}(10^{-11})$ vs. the much
smaller vacuum value ${{|\Delta\bar{c}_\theta|}\over{c}}\lesssim
10^{-15}$) and, at the same time, can help to understand the
differences and the analogies with experiments in strongly bound
solid dielectrics where the refractivity is ${\cal O}(1)$ but again
an instantaneous value ${{|\Delta\bar{c}_\theta|}\over{c}}\lesssim
10^{-15}$ (as in the vacuum case) is presently observed.

We emphasize that, due to this alternative view of the ether-drift
experiments in gases, precise measurements where optical cavities
are maintained in an extremely high vacuum, both at room temperature
and in the cryogenic regime, become essential to exclude a purely
thermal interpretations of light anisotropy, as for instance
ultimately associated with the CMB temperature dipole. In fact, once
any residual gaseous matter is totally negligible, the definite
persistence in vacuum of the $10^{-15}$ instantaneous signal would
support the more radical idea of a genuine preferred frame for
relativity.

In the end, Sect.8 will contain a summary and our conclusions.

\section{A modern version of Maxwell's calculation}

To start with, let us introduce the parameter $\epsilon$ defined by
the relation $\epsilon=({\cal N} -1) \ll 1$, ${\cal N}$ being the
refractive index of the medium where light propagates. For instance,
the medium (e.g. a gas) could fill an optical cavity at rest in a
frame $S'$ which moves with uniform velocity $v$ with respect to the
hypothetical $\Sigma$. Now, by assuming i) that the velocity of
light is exactly isotropic when $S'\equiv \Sigma$ and ii) the
validity of Lorentz transformations, then any anisotropy in $S'$
should vanish identically  either for $v = 0$ or for the ideal
vacuum case ${\cal N} = 1$ when the velocity of light $c_\gamma$
coincides with the basic parameter $c$ entering Lorentz
transformations \footnote{Actually, Guerra and De Abreu have shown
\cite{guerra2005} that the null result of a Michelson-Morley
experiment in an ideal vacuum can also be deduced without using
Lorentz transformations, but only from general assumptions on the
choice of the admissible clocks.}. Thus, one can expand in powers of
the two small parameters $\epsilon$ and $\beta=v/c$. By taking into
account that, by its very definition, the two-way velocity
$\bar{c}_\gamma(\theta)$ is invariant under the replacement $\beta
\to -\beta$ and that, for any fixed $\beta$, is also invariant under
the replacement $\theta \to \pi +\theta$, to lowest non-trivial
level ${\cal O}(\epsilon\beta^2)$, one finds the general expression
\cite{plus,foop}
\begin{eqnarray}
\label{legendre}
       \bar{c}_\gamma(\theta) \sim {{c}\over{ {\cal N} }} \left[1- \epsilon~\beta^2
\sum^\infty_{n=0}\zeta_{2n}P_{2n}(\cos\theta)
  \right]
\end{eqnarray}
Here, to take into account invariance under $\theta \to \pi
+\theta$, the angular dependence has been given as an infinite
expansion of even-order Legendre polynomials with arbitrary
coefficients $\zeta_{2n}={\cal O}(1)$. In Einstein's special
relativity, where there is no preferred reference frame, these
$\zeta_{2n}$ coefficients should vanish identically. In a
``Lorentzian'' approach, on the other hand, there is no reason why
they should vanish {\it a priori}.

By leaving out the first few $\zeta$'s as free parameters in the
fits, Eq.(\ref{legendre}) can already represent a viable form to
compare with experiments. Still, one can further sharpen the
predictions by exploiting one more derivation of the $\epsilon \to
0$ limit with a preferred frame. This other argument is based on the
effective space-time metric $g^{\mu\nu}=g^{\mu\nu}({\cal N})$ which,
through the relation $g^{\mu\nu}p_\mu p_\nu=0$, describes light
propagation in a medium of refractive index ${\cal N}$, see e.g.
\cite{leonhardt} and references quoted therein. For the quantum
theory, a derivation of this metric from first principles was given
by Jauch and Watson \cite{jauch} who worked out the quantization of
the electromagnetic field in a dielectric. They noticed that the
procedure introduces unavoidably a preferred reference frame, the
one where the photon energy spectrum does not depend on the
direction of propagation, and which is ``usually taken as the system
for which the medium is at rest''. However, such an identification
reflects the point of view of special relativity with no preferred
frame. Instead, one can adapt their results to the case where the
angle-independence of the photon energy is only valid when both
medium and observer are at rest in some particular frame $\Sigma$.

With this premise, let us consider two identical optical cavities,
namely cavity 1, at rest in $\Sigma$, and cavity 2, at rest in $S'$,
and denote by $\pi_\mu\equiv ( {{E_\pi}\over{c}},{\bf \pi}) $ the
light 4-momentum for $\Sigma$ in his cavity 1 and by $p_\mu\equiv (
{{E_p}\over{c}},{\bf p})$ the corresponding light 4-momentum for
$S'$ in his cavity 2. Let us also denote by $g^{\mu\nu}$ the
space-time metric that $S'$ uses in the relation $g^{\mu\nu}p_\mu
p_\nu=0$ and by
\begin{equation}
\label{metricsigma}\gamma^{\mu\nu}={\rm diag}({\cal N}^2,-1,-1,-1)
\end{equation}
the metric used by $\Sigma$ in the relation
$\gamma^{\mu\nu}\pi_\mu\pi_\nu=0$ and which gives an isotropic
velocity $c_\gamma=E_\pi/|{\bf \pi}|={{c}\over{{\cal N}}}$. Notice
that, in this framework, special relativity is included as a
particular case where there is no observable difference between
$\Sigma$ and $S'$ and the two frames are placed on the same footing.

Let us first consider the ideal vacuum limit ${\cal N}=1$. Here, the
frame independence of the velocity of light requires to impose
\begin{equation} \label{limitingintro} g^{\mu\nu}({\cal N}=1)=
\gamma^{\mu\nu}({\cal N}=1)=\eta^{\mu\nu}\end{equation} where
$\eta^{\mu\nu}$ is the Minkowski tensor. This standard equality
amounts to introduce a transformation matrix, say $A^{\mu}_{\nu}$,
such that
\begin{equation}
\label{vacuum}
g^{\mu\nu}=A^{\mu}_{\rho}A^{\nu}_{\sigma}\eta^{\rho\sigma}=\eta^{\mu\nu}
\end{equation}
This relation is strictly valid for ${\cal N}=1$. However, by
continuity, one is driven to conclude that an analogous relation
between $g^{\mu\nu}$ and $\gamma^{\mu\nu}$ should also hold in the
$\epsilon \to 0$ limit. The only subtlety is that relation
(\ref{vacuum}) does not fix uniquely $A^{\mu}_{\nu}$. In fact, one
can either choose the identity matrix, i.e.
$A^{\mu}_{\nu}=\delta^{\mu}_{\nu}$, or a Lorentz transformation,
i.e. $A^{\mu}_{\nu}=\Lambda^{\mu}_{\nu}$. Since for any finite $v$
these two matrices cannot be related by an infinitesimal
transformation, it follows that $A^{\mu}_{\nu}$ is a two-valued
function in the $\epsilon \to 0$ limit.

Therefore, in principle, there are two solutions. Namely, if
$A^{\mu}_{\nu}$ is the identity matrix, we expect a first solution
\begin{equation}\left[g^{\mu\nu}({\cal N})\right]_1=
\gamma^{\mu\nu}\sim \eta^{\mu\nu} + 2\epsilon \delta^\mu_0
\delta^\nu_0\end{equation} while, if $A^{\mu}_{\nu}$ is a Lorentz
transformation, we expect the other solution
\begin{equation} \label{2intro} \left[g^{\mu\nu}({\cal
N})\right]_2= \Lambda^{\mu}_{\rho}
\Lambda^{\nu}_{\sigma}\gamma^{\rho\sigma} \sim \eta^{\mu\nu} +
2\epsilon v^\mu v^\nu
\end{equation} $v^\mu$ being the dimensionless
$S'$ 4-velocity, $v^\mu\equiv(v^0,{\bf v}/c)$ with $v_\mu v^\mu=1$.

Notice that with the former choice, implicitly adopted in special
relativity to preserve isotropy in all reference systems also for
${\cal N} \neq 1$, one is introducing a discontinuity in the
transformation matrix for any $\epsilon \neq 0$. Indeed, the whole
emphasis on Lorentz transformations depends on enforcing
Eq.(\ref{vacuum}) for $A^{\mu}_{\nu}=\Lambda^{\mu}_{\nu}$ so that
$\Lambda^{\mu \sigma}\Lambda^{\nu}_{\sigma}=\eta^{\mu\nu}$ and the
Minkowski metric applies to all equivalent frames.

On the other hand, with the latter solution, by replacing in the
relation $p_\mu p_\nu g^{\mu\nu}=0$, the photon energy now depends
on the direction of propagation. Then, by defining the light
velocity $c_\gamma(\theta)$ from the ratio $E_p/|{\bf p}|$, one
finds \cite{plus,foop} \BE \label{oneway0}
       c_\gamma(\theta) \sim {{c} \over{{\cal N}}}~\left[
       1- 2\epsilon \beta \cos\theta -
       \epsilon \beta^2(2-\sin^2\theta)\right]
\EE and a two-way velocity
\begin{eqnarray}\label{3intro}
       \bar{c}_\gamma(\theta)={{2~c_\gamma(\theta)c_\gamma(\pi +\theta)}\over{c_\gamma(\theta)
+c_\gamma (\pi +\theta)}}
       &\sim& (c/ {\cal N})\left[1-\epsilon \beta^2\left(2 -
       \sin^2\theta\right) \right]
\end{eqnarray}
where $\theta$ is the angle between ${\bf v}$ and $\bf p$ (as
defined in the $S'$ frame).

Eq.(\ref{3intro}) corresponds to setting in Eq.(\ref{legendre})
$\zeta_0=4/3$, $\zeta_{2}= 2/3$ and all $\zeta_{2n}=0$ for $n
> 1$ and can be considered a modern version of Maxwell's original
calculation. It represents a definite, alternative model for the
interpretation of experiments performed close to the ideal vacuum
limit $\epsilon = 0$, such as in gaseous systems, and will be
adopted in the following.

A conceptual detail concerns the relation of the gas refractive
index ${\cal N}$, as introduced in Eq.(\ref{metricsigma}), to the
experimental quantity ${\cal N}_{\rm exp}$ which is extracted from
measurements of the two-way velocity in the earth laboratory. By
introducing a $\theta-$dependent refractive index through the
relation
\begin{equation}
\label{refractivetheta} {\bar c_\gamma (\theta)}\equiv
{{c}\over{\bar{\cal N}(\theta)}}
\end{equation}
one should thus define the experimental value by an angular average
of Eq.(\ref{3intro}), i.e.
\begin{equation} \label{nexp} {{c}\over{ {\cal N}_{\rm exp} }}\equiv
\langle {{c}\over{\bar{\cal N}(\theta)}} \rangle_\theta= {{c}\over{
{\cal N}  }} ~\left[1-{{3}\over{2}} ({\cal N} -1)\beta^2\right]
\end{equation}
From this relation, one can determine in principle the unknown value
${\cal N}  \equiv {\cal N}(\Sigma)$ (as if the container of the gas
were at rest in $\Sigma$), in terms of the experimentally known
quantity ${\cal N}_{\rm exp}\equiv{\cal N}(earth)$ and of $v$. For
instance, for air the most precise determinations are at the level
$10^{-7}$, say ${\cal N}_{\rm exp}=1.0002924..$ for light of 589 nm,
at 0 $^o$C and atmospheric pressure. In practice, for the standard
velocity values involved in most cosmic motions, say $ v \sim $ 300
km/s, the difference between ${\cal N}(\Sigma)$ and ${\cal
N}(earth)$ is below $10^{-9}$ and thus completely negligible. The
same holds true for the other gaseous systems (say nitrogen, carbon
dioxide, helium,..) for which the present experimental accuracy in
the refractive index is, at best, at the level $10^{-7}$. Finally,
the isotropic two-way speed of light is better determined in the
low-pressure limit where $({\cal N}-1)\to 0$. In the same limit, for
any given value of $v$, the approximation ${\cal N}(\Sigma)={\cal
N}(earth)$ becomes better and better.

From Eq.(\ref{3intro}) we obtain a fractional anisotropy
\begin{equation} \label{bbasic2new} {{\Delta \bar{c}_\theta }
\over{c}}
={{\bar{c}_\gamma(\pi/2+\theta)-\bar{c}_\gamma(\theta)}\over{c}}\sim
     \epsilon~
       {{v^2 }\over{c^2}} \cos2(\theta-\theta_0) \end{equation}
Here $v$ and $\theta_0$ are respectively the magnitude and the
direction of the drift in the plane of the interferometer so that,
from Eq.(\ref{newintro}), one finds directly the fringe pattern
\begin{equation} \label{newintro1} {{\Delta
\lambda(\theta)}\over{\lambda}}= {{2D}\over{\lambda}} ~{{\Delta
\bar{c}_\theta } \over{c}}\sim 2\epsilon~
{{D}\over{\lambda}}{{v^2}\over{c^2}}\cos 2(\theta -\theta_0)
\end{equation}  In this scheme, the ether drift is a
second-harmonic effect, i.e. periodic in the range $[0,\pi]$, as in
the classical theory (see e.g. \cite{kennedy} for a simple
derivation). Only its amplitude
 \begin{equation} \label{a2}
 A_2=2\epsilon~{{ D }\over{\lambda}}{{{v}^2}\over{c^2}}
\end{equation}
is suppressed by the very small factor $2\epsilon$ with respect to
the classical prediction
\begin{equation} \label{a2class}
 A^{\rm class}_2={{ D }\over{\lambda}}{{{v}^2}\over{c^2}}
\end{equation}
Thus one can re-absorb all effects into a much smaller {\it
observable} velocity
\begin{equation}  \label{vobs} v^2_{\rm obs} \sim 2\epsilon v^2 \end{equation}
which depends on the gas refractive index and is the one
traditionally reported in the classical analysis of the data.

However, as anticipated in the Introduction, for a proper comparison
with experiments a change of perspective is needed in the physical
description of the ether-drift phenomenon. In the following section,
we will illustrate a simple stochastic model that we propose for the
analysis of the data.

\section{A stochastic form of ether-drift}

To make explicit the time dependence of the signal let us re-write
Eq.(\ref{bbasic2new}) as \begin{equation} \label{basic2}
     {{\Delta \bar{c}_\theta(t) } \over{c}}
    \sim
 \epsilon {{v^2(t) }\over{c^2}}\cos 2(\theta
-\theta_0(t)) \end{equation} where $v(t)$ and $\theta_0(t)$ indicate
respectively the instantaneous magnitude and direction of the drift
in the plane of the interferometer. This can also be re-written as
\begin{equation} \label{basic3} {{\Delta \bar{c}_\theta(t) } \over{c}}\sim
2{S}(t)\sin 2\theta +
      2{C}(t)\cos 2\theta \end{equation} with \begin{equation} \label{amplitude10}
       2C(t)= \epsilon~ {{v^2_x(t)- v^2_y(t)  }
       \over{c^2}}~~~~~~~2S(t)=\epsilon ~{{2v_x(t)v_y(t)  }\over{c^2}}
\end{equation} and $v_x(t)=v(t)\cos\theta_0(t)$, $v_y(t)=v(t)\sin\theta_0(t)$

As anticipated in the Introduction, the standard assumption to
analyze the data is based on the idea of smooth, regular modulations
of the signal associated with a cosmic earth velocity. In general,
this is characterized by a magnitude $V$, a right ascension $\alpha$
and an angular declination $\gamma$. These parameters can be
considered constant for short-time observations of a few days where
there are no appreciable changes due to the earth orbital velocity
around the sun. In this framework, where the only time dependence is
due to the earth rotation, the traditional identifications are
$v(t)\equiv \tilde v(t)$ and $\theta_0(t)\equiv\tilde\theta_0(t)$
where $\tilde v(t)$ and $\tilde\theta_0(t)$ derive from the simple
application of spherical trigonometry \cite{nassau}
\begin{equation} \label{nassau1}
       \cos z(t)= \sin\gamma\sin \phi + \cos\gamma
       \cos\phi \cos(\tau-\alpha)
\end{equation} \begin{equation} \label{projection}
       \tilde {v}(t) =V \sin z(t)
\end{equation} \begin{equation} \label{nassau2}
    \tilde{v}_x(t) = \tilde{v}(t)\cos\tilde\theta_0(t)= V\left[ \sin\gamma\cos \phi -\cos\gamma
       \sin\phi \cos(\tau-\alpha)\right]
\end{equation} \begin{equation} \label{nassau3}
      \tilde{v}_y(t)= \tilde{v}(t)\sin\tilde\theta_0(t)= V\cos\gamma\sin(\tau-\alpha) \end{equation}
Here $z=z(t)$ is the zenithal distance of ${\bf{V}}$, $\phi$ is the
latitude of the laboratory, $\tau=\omega_{\rm sid}t$ is the sidereal
time of the observation in degrees ($\omega_{\rm sid}\sim
{{2\pi}\over{23^{h}56'}}$) and the angle $\theta_0$ is counted
conventionally from North through East so that North is $\theta_0=0$
and East is $\theta_0=90^o$. With the identifications $v(t)\equiv
\tilde v(t)$ and $\theta_0(t)\equiv\tilde\theta_0(t)$, one thus
arrives to the simple Fourier decomposition \begin{equation}
\label{amorse1}
      S(t)\equiv {\tilde S}(t) =S_0+
      {S}_{s1}\sin \tau +{S}_{c1} \cos \tau
       + {S}_{s2}\sin(2\tau) +{S}_{c2} \cos(2\tau)
\end{equation}
\begin{eqnarray}
 \label{amorse2}
       C(t)\equiv {\tilde C}(t)=
       {C}_0 + {C}_{s1}\sin \tau +{C}_{c1} \cos \tau
       + {C}_{s2}\sin(2 \tau) +{C}_{c2} \cos(2 \tau)
\end{eqnarray}
where the  $C_k$ and $S_k$ Fourier coefficients depend on the three
parameters $(V,\alpha,\gamma)$ (see \cite{applied}) and, to very
good approximation, should be time-independent for short-time
observations.

However, identifying the instantaneous quantities $v_x(t)$ and
$v_y(t)$, with their counterparts $\tilde{v}_x(t)$ and
$\tilde{v}_y(t)$ is equivalent to assume a form of regular, laminar
flow where global and local velocity fields coincide. Instead, as
anticipated in the Introduction, one may consider the alternative
model of a turbulent flow where the two sets of quantities are only
{\it indirectly} related. This picture is motivated by the idea of
the vacuum as a stochastic medium for which the local velocity field
becomes non-differentiable and the ordinary formulation in terms of
differential equations breaks down \cite{onsager,eyink}. Thus, one
has to adopt some other description, for instance a formulation in
terms of random Fourier series \cite{onsager,landau,fung}. In this
other approach, the parameters of the macroscopic motion are only
used to fix the typical boundaries for a microscopic velocity field
which has an intrinsic non-deterministic nature.

The simplest model, adopted in refs.\cite{plus,physica}, corresponds
to a turbulence which, at small scales, appears homogeneous and
isotropic \footnote{This picture reflects the basic Kolmogorov
theory \cite{kolmo} of a fluid with vanishingly small viscosity.}.
The analysis of Sect.2, can then be embodied in an effective
space-time metric for light propagation
\begin{equation} \label{random} \hat{g}^{\mu\nu}(t) \sim \eta^{\mu\nu} + 2
\epsilon \hat{v}^\mu(t) \hat{v}^\nu(t) \end{equation} where
$\hat{v}^\mu(t)$ is a random 4-velocity field which describes the
drift and whose boundaries depend on a smooth field
$\tilde{v}^\mu(t)$ determined by the average earth motion. If this
corresponds to the actual physical situation,it is easy to see why a
genuine stochastic signal can become consistent with average values
$(C_k)^{\rm avg} = (S_k)^{\rm avg} = 0$ obtained by fitting the data
with Eqs.(\ref{amorse1}) and (\ref{amorse2}).

Our intention is to simulate the two components of the velocity in
the x-y plane, at a given fixed location in the laboratory, to
reproduce the $S(t)$ and $C(t)$ functions Eq.(\ref{amplitude10}).
For a homogeneous turbulence, one finds the general expressions
\begin{equation} \label{vx} \hat{v}_x(t)= \sum^{\infty}_{n=1}\left[
       x_n(1)\cos \omega_n t + x_n(2)\sin \omega_n t \right] \end{equation}
\begin{equation} \label{vy} \hat{v}_y(t)= \sum^{\infty}_{n=1}\left[
       y_n(1)\cos \omega_n t + y_n(2)\sin \omega_n t \right] \end{equation}
where $\omega_n=2n\pi/T$, T being a time scale which represents a
common period of all stochastic components. For numerical
simulations, the typical value $T=T_{\rm day}$= 24 hours was adopted
\cite{plus,physica}. However, it was also checked with a few runs
that the statistical distributions of the various quantities do not
change substantially by varying $T$ in the rather wide range
$0.1~T_{\rm day}\leq T \leq 10~T_{\rm day}$.

The coefficients $x_n(i=1,2)$ and $y_n(i=1,2)$ are random variables
with zero mean and have the physical dimension of a velocity. In
general, we can denote by $[-d_x(t),d_x(t)]$ the range for
$x_n(i=1,2)$ and by $[-d_y(t),d_y(t)]$ the corresponding range for
$y_n(i=1,2)$. Statistical isotropy would require to impose $d_x(t)=
d_y(t)$. However, to illustrate the more general case, let us first
consider $d_x(t) \neq d_y(t)$. In terms of these boundaries, the
only non-vanishing (quadratic) statistical averages are
\begin{equation} \label{quadratic} \langle x^2_n(i=1,2)\rangle_{\rm
stat}={{d^2_x(t) }\over{3 ~n^{2\eta}}}~~~~~~~~~~~\langle
y^2_n(i=1,2)\rangle_{\rm stat}={{d^2_y(t) }\over{3 ~n^{2\eta}}}
\end{equation} in a uniform-probability model within the intervals
$[-d_x(t),d_x(t)]$ and $[-d_y(t),d_y(t)]$. Here, the exponent $\eta$
controls the power spectrum of the fluctuating components. For
numerical simulations, between the two values $\eta=5/6$ and
$\eta=1$ reported in ref.\cite{fung}, we have adopted $\eta=1$ which
corresponds to the Lagrangian description where the point of
measurement is a wandering material point in the fluid.

Finally, the connection with the earth cosmic motion is obtained by
identifying $d_x(t)=\tilde v_x(t)$ and $d_y(t)=\tilde v_y(t)$ as
given in Eqs. (\ref{nassau1})$-$(\ref{nassau3}). In this case, it is
natural to adopt the set $V=$ 370 km/s, $\alpha=168$ degrees,
$\gamma$= -7 degrees, which describes the average earth motion with
respect to the CMB.

If, however, we require statistical isotropy, the relation
\begin{equation} \label{correct} \tilde{v}^2_x(t) +
\tilde{v}^2_y(t)=\tilde{v}^2(t)\end{equation}  requires the
identification \footnote{The correct normalization
Eq.(\ref{correct}) produces boundaries which are smaller by a factor
${{1}\over{\sqrt{ 2}}}$ as compared to those of ref.\cite{physica}
where the relation $d_x(t)=d_y(t) \sim \tilde{v}(t)$ was assumed.
For this reason, in view of Eqs.(\ref{quadratic}), the resulting
 amplitudes of the signal are now predicted to be smaller by about a factor of 2.}
\begin{equation} \label{isot} d_x(t)=d_y(t)={{ \tilde{v}(t)}\over{\sqrt{2} }} \end{equation}
For such isotropic model, by combining Eqs.(\ref{vx})$-$(\ref{isot})
and in the limit of an infinite statistics, one gets
\begin{eqnarray}
\label{vanishing} \langle \hat{v}^2_x(t)\rangle_{\rm stat}=\langle
\hat{v}^2_y(t)\rangle_{\rm
stat}={{\tilde{v}^2(t)}\over{2}}~{{1}\over{3}} \sum^{\infty}_{n=1}
{{1}\over{n^2}} ={{\tilde{v}^2(t)}\over{2}}~
{{\pi^2}\over{18}}\nonumber \\ \langle
\hat{v}_x(t)\hat{v}_y(t)\rangle_{\rm stat}=0
\end{eqnarray}
and  vanishing statistical averages
\begin{equation} \label{vanishing2}\langle C(t)\rangle_ {\rm
stat}=0~~~~~~~~~~~~~~~~~~\langle S(t)\rangle_ {\rm stat}=0
\end{equation} at {\it any} time $t$, see Eqs.(\ref{amplitude10}).
Therefore, by construction, this model gives a definite non-zero
signal but, if the same signal were fitted with Eqs.(\ref{amorse1})
and (\ref{amorse2}), it also gives average values $(C_k)^{\rm
avg}=0$, $(S_k)^{\rm avg}=0$ for the Fourier coefficients.

\section{The classical experiments in gaseous media}

To fully appreciate the change of perspective implied by
Eqs.(\ref{vanishing2}), let us consider the traditional procedure of
data taking in the classical experiments. Fringe shifts were
observed at the same sidereal time  on a few consecutive days so
that changes in the earth orbital  velocity could be ignored. Then,
see Eqs.(\ref{newintro1}) and (\ref{basic3}), the data were averaged
at any given angle $\theta$
\begin{equation} \label{averagefringe}\langle{{\Delta
\lambda(\theta;t)}\over{\lambda}}\rangle_ {\rm stat}=
{{2L}\over{\lambda}} \left[2\sin 2\theta~\langle S(t)\rangle_ {\rm
stat} + 2\cos 2\theta~\langle C(t)\rangle_ {\rm stat} \right]
\end{equation} and these averages were compared with various models
of cosmic motion.

But, if the ether-drift is a genuine stochastic phenomenon, as
expected if the physical vacuum were similar to a turbulent fluid
which becomes isotropic at small scales, these average combinations
should vanish {\it exactly} for an infinite number of measurements.
Thus, averages of vectorial quantities are non vanishing just
because the statistics is finite and forming the averages
Eq.(\ref{averagefringe}) is not a meaningful procedure. In
particular, the direction $\theta_0(t)$ of the drift in the plane of
the interferometer (defined by the relation $\tan2\theta_0(t)=
S(t)/C(t)$) is a completely random quantity which has no definite
limit by combining a large number of observations.

Instead, one should concentrate on the 2nd-harmonic amplitudes by
comparing with \begin{equation} \label{AA}
A_2(t)={{2L}\over{\lambda}}~ 2\sqrt{S^2(t) + C^2(t)} \end{equation}
These are positive-definite quantities and, as such, remain
definitely non-zero after any averaging procedure. In addition,
being rotationally invariant, their statistical properties remain
the same by adopting the isotropic model Eq.(\ref{isot}) or the
non-isotropic choice $d_x(t)\equiv \tilde v_x(t)$ and $d_y(t)\equiv
\tilde v_y(t)$.

As a matter of fact, by restricting to the amplitudes, one finds
\cite{plus} a good consistency of the data with the kinematical
parameters obtained from the CMB observations. For instance, let us
consider the experimental 2nd-harmonics extracted from the six
sessions of the Michelson-Morley experiment, see Table 1.

\begin{table*}
\caption{\it The amplitude of the fitted second-harmonic component
$A^{\rm EXP}_2$ for the six experimental sessions of the
Michelson-Morley experiment. The table is taken from
ref.\cite{plus}.}
\begin{center}
\begin{tabular}{cl}
\hline
SESSION       & ~~~~~~      $A^{\rm EXP}_2$   \\
\hline
July 8  (noon) & $0.010 \pm 0.005$  \\
July 9  (noon) & $0.015 \pm 0.005$   \\
July 11 (noon) & $0.025 \pm 0.005$    \\
July 8  (evening) & $0.014 \pm 0.005$  \\
July 9  (evening) &$0.011 \pm 0.005$   \\
July 12 (evening) & $0.024 \pm 0.005$  \\
\hline

\end{tabular}
\end{center}
\end{table*}

\begin{figure}
\begin{center}
\includegraphics[scale=0.90]{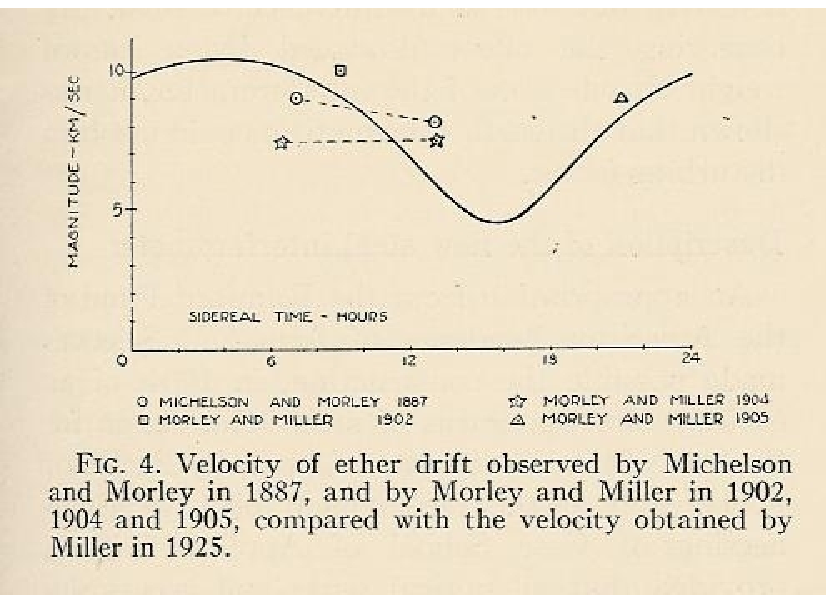}
\end{center}
\caption{\it The magnitude of the observable velocity measured in
various experiments as reported by Miller
\cite{miller}.}\label{miller}
\end{figure}

Due to their reasonable statistical consistency, one can compute the
mean and variance of the six determinations by obtaining $A^{\rm
EXP}_2 \sim 0.016 \pm 0.006$. By comparing with the classical
prediction $A^{\rm class}_2={{L}\over{\lambda}} {{(30 {\rm km/s})^2
}\over{c^2}}\sim 0.20$, this average amplitude corresponds to  an
observable velocity $ v_{\rm obs} \sim (8.4\pm 1.6) $ km/s in very
good agreement with Miller's analysis, see Fig.\ref{miller}. Then,
by using Eq.(\ref{vobs}), for air at atmospheric pressure where
$\epsilon \sim 2.8\cdot 10^{-4}$, one obtains a true kinematical
value $ v \sim (355 \pm 70)$ km/s. Notice the consistency with the
determination $  v \sim$ 370 km/s obtained from the CMB
observations.

Analogously, let us consider Miller's extensive observations. After
the critical re-analysis of his original measurements performed by
the Shankland team \cite{shankland}, one has an accurate
determination of the overall average for all epochs of the year (see
Table III of \cite{shankland}). The resulting amplitude $A^{\rm
EXP}_2=0.044 \pm 0.022$, when compared with the equivalent classical
prediction for Miller's interferometer $A^{\rm
class}_2={{L}\over{\lambda}} {{(30 {\rm km/s})^2 }\over{c^2}}\sim
0.56$, gives $v_{\rm obs}\sim (8.4\pm 2.2)$  km/s and, by using
Eq.(\ref{vobs}), a true kinematical value $v \sim (355\pm 90)$ km/s,
again consistent with the CMB observations.

At the same time, the perfect identity of two determinations
obtained in completely different experimental conditions (in the
basement of Cleveland University or on top of Mount Wilson)
indicates that the standard interpretation \cite{shankland} of the
residuals in terms of temperature differences in the air of the
optical paths is only acceptable provided this gradient has a {\it
non-local} origin. A natural physical interpretation will be
proposed in Section 7.

Analogous considerations can be applied to the other classical
experiments in gaseous helium, such as Illingworth's 1927 experiment
at Caltech (sensitivity about 1/1500 of a fringe) or  Joos's 1930
experiment in Jena (sensitivity about 1/3000 of a fringe). By
ignoring the directional character of the data and just restricting
to the amplitudes of the individual observations \cite{plus}, for
$\epsilon \sim 3.3\cdot 10^{-5}$ in Eq.(\ref{vobs}), the very low
observable velocities of about 2$\div$3 km/s become consistent with
the CMB value of 370 km/s. In particular, by using Eqs.(\ref{AA})
and (\ref{projection}) to fit the time dependence of the amplitudes
extracted from Joos's observations (data collected at regular steps
of one hour to cover the full sidereal day and recorded
automatically by photocamera), one even gets \cite{plus} some
information on the right ascension and angular declination, namely
$\alpha({\rm fit-Joos})= (168 \pm 30)$ degrees and $\gamma({\rm
fit-Joos})= (-13 \pm 14)$ degrees, to compare with the present
values $\alpha({\rm CMB}) \sim$ 168 degrees and $\gamma({\rm CMB})
\sim -$7 degrees.

The only possible discrepancy found in ref.\cite{plus} concerned the
Michelson-Pease-Pearson (MPP) experiment at Mount Wilson which was
giving a considerably smaller central value, namely $v \sim$ 180
km/s, for the kinematical velocity, even though the associated
uncertainty could not be estimated. The general situation is
summarized in Table 2 where we have also included the determinations
from the Tomaschek \cite{tomaschek1} and Piccard-Stahel
\footnote{The velocities for the Piccard-Stahel experiment
\cite{piccard3} derive from the value $D/\lambda=6.4\cdot 10^6$ and
the average 2nd-harmonic amplitude $(2.8 \pm 1.5)\cdot 10^{-3}$.
This is obtained from their individual 24 determinations namely (in
units $10^{-3}$), the 12 Mt.Rigi values $A^{\rm EXP}_2=$ 3.4, 1.1,
4.0, 2.4, 2.4, 4.3, 2.3, 2.6, 0.6, 2.0, 1.2, 3.9, and the 12
Brussels measurements, at night $A^{\rm EXP}_2=$ 3.2, 5.2, 6.5, 2.2,
4.9, 3.8 and in the morning $A^{\rm EXP}_2=$ 1.85, 1.27, 3.40, 1.00,
3.70, 1.14.} experiments \cite{piccard3}. However, we will now show
that, within statistical uncertainties, also the MPP experiment can
become consistent with our stochastic model.

\begin{table*}
\caption {\it The average velocity observed (or the limits placed)
by the classical ether-drift experiments in the alternative
interpretation where the relation between the observable $v_{\rm
obs}$ and the kinematical $v$ is governed by Eq.(\ref{vobs}).}
\begin{center}
\begin{tabular}{clll}
\hline Experiment &gas in the interferometer
&~~~~$v_{\rm obs}({\rm km/s})$ & ~~~~$v$({\rm km/s})\\
\hline
Michelson-Morley(1887)   & ~~~~~~~~~~ air & ~~~~ $8.4^{+1.5}_{-1.7}$&~~~$355^{+62}_{-70}$ \\
Morley-Miller(1902-1905)  & ~ ~~~~~~~~~air& ~~~~ $8.5\pm 1.5$ & ~~ $ 359 \pm 62$ \\
Miller(1925-1926) & ~~~~~~~~~~~air & ~~~~~$8.4^{+1.9}_{-2.5}$ &~~~$355^{+79}_{-104}$ \\
Tomaschek (1924)&~~~~~~~~~~~air& ~~~~~$7.7^{+2.1}_{-2.8}$&~~~$325^{+87}_{-116}$  \\
Kennedy(1926)  & ~~~~~~~~~~~helium &~~~~~~$<5 $  &  ~~~$<600 $\\
Illingworth(1927) &~~~~~~~~~~~helium  &~~~~~$2.4^{+0.8}_{-1.2}$ &~~~$295^{+98}_{-146}$\\
Piccard-Stahel(1926-1927)&~~~~~~~~~~~air     &~~~~~$6.3^{+1.5}_{-2.0}$ &~~~$266^{+62}_{-83}$   \\
Michelson-Pease-Pearson(1929)& ~~~~~~~~~~~air &~~~~~$4.3\pm... $  &  ~~~$ 182 \pm ... $\\
Joos(1930)  &~~~~~~~~~~~helium&~~~~$~ 1.8^{+0.5}_{-0.6} $  & ~~~$226^{+63}_{-76} $\\
\hline
\end{tabular}
\end{center}
\end{table*}

As discussed in \cite{plus} it is extremely difficult to understand
the results of the MPP experiment from the original articles
\cite{mpp,mpp2}. No numerical results are reported and the two
papers are even in contradiction about the magnitude of the measured
effects (``one-fifteenth'' of the expected value vs.
``one-fiftieth''). To try to understand, we have consulted another
article which, rather surprisingly, was signed by Pease alone
\cite{pease}. In this article, Pease declares that, in their
experiment, to test Miller's claims, they concentrated on a purely
{\it differential } type of measurement. For this reason, he only
reports the quantity \[ \delta (\theta)= \langle
{{\Delta\lambda(\theta;t=5:30) }\over{\lambda}} \rangle_{\rm stat} -
\langle {{\Delta\lambda(\theta;t=17:30) }\over{\lambda}}
\rangle_{\rm stat} \] This means that they were performing a large
set of observations at sidereal time 5:30 and averaging the data.
Then, the same procedure was carried out, in the same days, at
sidereal time 17:30. Finally, the two averages were subtracted to
form the quantities $\delta(\theta)$. These are typically below $\pm
0.004$ and this is the order of magnitude which is usually compared
\cite{shankland} with the classical expectation for the MPP
apparatus, namely $A^{\rm class}_2={{L}\over{\lambda}} {{(30 {\rm
km/s})^2 }\over{c^2}}\sim 0.45$ for optical path of 85 feet or
$A^{\rm class}_2={{L}\over{\lambda}} {{(30 {\rm km/s})^2
}\over{c^2}}\sim 0.29$ for optical path of 55 feet.

\begin{figure}
\begin{center}
\includegraphics[scale=0.45]{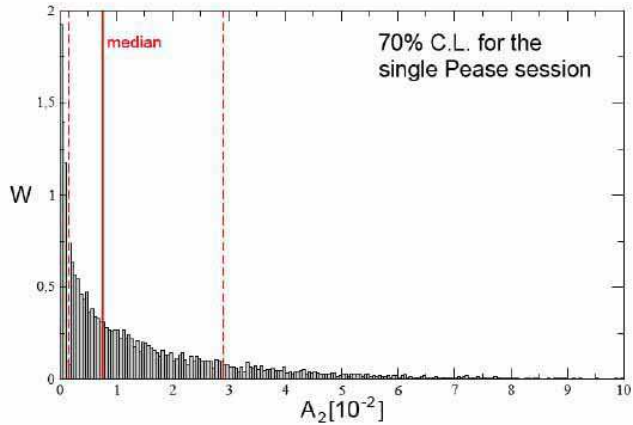}
\end{center}
\caption{ {\it The histogram W of a numerical simulation of 10.000
instantaneous amplitudes for the single session of January 13, 1928,
reported by Pease \cite{pease}. The vertical normalization is to a
unit area. We show the median and the 70$\%$ CL. The limits on the
random Fourier components in Eqs.(\ref{vx}) and (\ref{vy}) were
fixed by the CMB kinematical parameters as explained in the text.}}
\label{pease}
\end{figure}

As explained above, by accepting a stochastic picture of the
ether-drift, the vector average of more and more observations will
wash out completely the physical information contained in the
original measurements. Therefore, from these $\delta-$values,
nothing can be said about the magnitude of the fringe shifts $
{{\Delta\lambda(\theta) }\over{\lambda}}$ obtained in the individual
measurements, i.e. before any averaging procedure and before any
subtraction. Pease just reports a poor-quality plot of a single
observation, performed on January 13, 1928, when the length of the
optical path was still 55 feet. In this plot, the fringes vary
approximately in the range $\pm 0.006$ whose absolute value may be
taken to estimate the amplitude of that observation.

We have thus performed a numerical simulation in our stochastic
model by generating 10,000 values of the amplitude, at the same
sidereal time 5:30 of the observation reported by Pease, and using
the CMB kinematical parameters to bound the random Fourier
components of the velocity field Eqs.(\ref{vx}) and (\ref{vy}). The
resulting histogram, reported in Fig.\ref{pease}, shows that the
value $A_2 \sim 0.006$ lies well within the 70$\%$ Confidence Limit.
Notice the large probability content at very small amplitudes
\footnote{Strictly speaking, for a more precise comparison with the
data, one should fold the histogram with a smearing function which
takes into account the finite resolution $\Delta$ of the apparatus.
The resulting curve will bend for $A_2 \to 0$ and saturate to a
limit which depends on $\Delta$. Nevertheless, this refinement
should not modify substantially the probability content around the
median which is very close to $A_2= 0.007$.} and the long tail
extending up to $A_2=0.030$ or even larger values.

The wide interval of amplitudes corresponding to the 70$\%$ C. L.
(which could be expressed as $0.014^{+0.015}_{-0.012}$) indicates
that, in our stochastic model, one could accomodate individual MPP
observations with an amplitude as 0.002 or as 0.030 which is fifteen
times larger. This is another crucial difference with a
deterministic model of the ether-drift. In this traditional view, in
fact, the amplitude can vary at most by a factor $r=(v_{\rm
max}/v_{\rm min})^2$ where $v_{\rm max}$ and $v_{\rm min}$ are
respectively the maximum and minimum daily projection of the earth
velocity in the interferometer plane. Therefore, since $r$ varies
typically by a factor of two, the observation of such large
fluctuations in the data would induce to conclude, in a
deterministic model, that there must be some systematic effect which
modifies the measurements in an uncontrolled way.

\begin{figure}
\begin{center}
\epsfig{figure=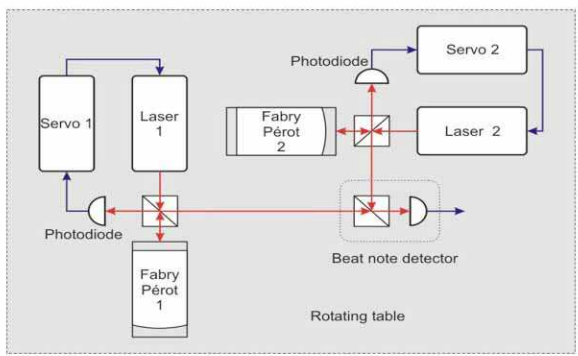,width=8truecm,angle=0}
\end{center}
\caption{ {\it The scheme of a modern ether-drift experiment. The
light frequencies are first stabilized by coupling the lasers to
Fabry-Perot optical resonators. The frequencies $\nu_1$ and $\nu_2$
of the signals from the resonators are then compared in the beat
note detector which provides the frequency shift $\Delta \nu=\nu_1
-\nu_2$. In present experiments a very high vacuum is maintained
within the resonators. } } \label{Fig.apparatus}
\end{figure}

This confirms the overall consistency of our picture with the
classical experiments and should induce to perform the new dedicated
experiments where the optical resonators which are coupled to the
lasers (see Fig.\ref{Fig.apparatus}) are filled by gaseous media. In
this case, from Eq.(\ref{bbasic2}), one should compare the data with
the prediction
\begin{equation} \label{bbbasic2}
 {{\Delta \nu(\theta) }\over{\nu_0}}  =
      {{\Delta \bar{c}_\theta } \over{c}} \sim \epsilon~
       {{v^2 }\over{c^2}} \cos2(\theta-\theta_0) \end{equation}

These experiments will likely require a good deal of ingenuity and
technical skill. For instance, an important element to increase the
overall stability and minimize systematic effects may consist in
obtaining the two optical resonators from the same block of material
as with the crossed optical cavity of ref.\cite{crossed}. Still,
measuring precisely the frequency shift in the gas mode will be a
delicate issue. To fix the ideas, let us consider gaseous helium at
atmospheric pressure, a velocity $v=300$ km/s and a typical laser
frequency of about $3\cdot 10^{14}$ Hz. In these conditions, the
expected shift is $\Delta \nu \sim 10$ kHz. This is much smaller
than many effects which must preliminarily be subtracted. For
instance, by changing from vacuum to the gas case under pressure,
and for a typical cavity length of 10 cm, the effect of cavity
deformations is about 10 MHz \cite{stone}. Theoretically, this
should not depend on the gas used but only on the solid parts of the
apparatus. Yet, experimental measurements at atmospheric pressure
show that there is a difference between Nitrogen and Helium of about
0.6 MHz \cite{stone}. Therefore, one should lower the pressure to
reduce this spurious effect. Of course, this is what might show up
in a single cavity while we are interested in the frequency shift
between two cavities where the effect will be reduced. Nevertheless,
the pressure will have to be lowered and, then, also the signal will
be reduced correspondingly. Therefore, several technical problems
must be solved before concluding that, in the gas case, there is a
definite improvement with respect to the classical experiments (in
particular with respect to Joos).

However, at present, a first rough check of Eq.(\ref{bbbasic2}) can
be obtained from the time variation of the signal observed in the
only modern experiment performed in similar conditions, namely the
1963 MIT experiment by Jaseja et. al \cite{jaseja} with He-Ne
lasers. At that time, the laser stabilization mechanism had not yet
been invented and one was just comparing directly the frequencies of
the two lasers. As a matter of fact, for a laser frequency
$\nu_0\sim 2.6 \cdot 10^{14}$ Hz, the residual observed variations
of a few kHz are consistent with the refractive index ${\cal N}_{\rm
He-Ne}\sim 1.00004$ and the typical change of the earth cosmic
velocity at the latitude of Boston. For more details, see the
discussion given in \cite{epl}.

Meanwhile, waiting for the new dedicated experiments, one can try to
have a different check with vacuum experiments. The point is that,
as illustrated in the next section, for the physical vacuum the
equality ${\cal N}_v =1$ might not be exact.

\section{An effective refractivity for the physical vacuum}

The idea of an effective refractivity for the physical vacuum
becomes natural by adopting a different view of the curvature
effects observed in a gravitational field.

The usual perspective, derived from General Relativity, is that
these effects require the introduction of a non-trivial metric field
$g_{\mu\nu}(x)$ viewed as a fundamental modification of Minkowski
space-time. By {\it fundamental}, we mean that deviations from flat
space might also occur at extremely small scales, in principle
comparable to the Planck length. Though, it is an experimental fact
that many physical systems for which, at a fundamental level,
space-time is exactly flat are nevertheless described by an
effective curved metric in their hydrodynamic limit, i.e. at length
scales that are much larger than the size of their elementary
constituents.

For this reason several authors, see e.g.
\cite{barcelo1,barcelo2,volo,bosegravity}, have started to explore
those gravity-analogs (moving fluids, condensed matter systems with
a refractive index, Bose-Einstein condensates,...) which are known
in flat space. The ultimate goal is that, as with the deflection of
light in Euclidean space when propagating in a medium of variable
density, one might succeed to explain the curvature effects in a
gravitational field in terms of the hydrodynamic excitations of an
underlying form  of (quantum) ether.

We believe that there is a value in this attempt. In fact, beyond
the simple level of an analogy, there might be a deeper significance
if the properties of the underlying medium could be matched with
those of the physical vacuum of electroweak and strong interactions.
In this case, the so called vacuum condensates, which play a crucial
role for fundamental phenomena such as mass generation and quark
confinement, could also represent a bridge between gravity and
particle physics \cite{ultraweak}.

To be more definite, suppose that gravity originates from some
long-range fields $s_k(x)$. By this we mean that their typical
wavelengths are larger than some minimal scale (consistently with
the experimental verifications \cite{eotwash} of the 1/r law) and
that the deviation of the effective $g_{\mu\nu}(x)$ from the
Minkowski tensor $\eta_{\mu\nu}$ can be expressed as
\begin{equation} g_{\mu\nu}(x)-\eta_{\mu\nu}=\delta g_{\mu\nu}[s_k(x)]
\end{equation}
with $\delta g_{\mu\nu}[s_k=0]=0$. In this type of approach, as in
the original Yilmaz derivation \cite{yilmaz}, Einstein's equations
for the metric should be considered as algebraic identities which
follow directly from the equations of motion for the $s_k$'s in flat
space, after introducing a suitable stress tensor for these
inducing-gravity fields \footnote{In the simplest, original Yilmaz
approach \cite{yilmaz} there is only one inducing-gravity field
$s_0(x)$ which plays the role of the Newtonian potential.
Introducing its stress tensor $t^\mu_\nu(s_0)= -\partial^\mu
s_0\partial_\nu s_0 + 1/2\delta^\mu_\nu~\partial^\alpha
s_0\partial_\alpha s_0$, to match the Einstein tensor, produces
differences from the Schwarzschild metric which are beyond the
present experimental accuracy, see \cite{tupper}. }. In this way,
one could (partially) fill the conceptual gap with classical General
Relativity. As an immediate consequence, if the $s_k$'s represent
{\it excitations} of the physical vacuum, which therefore vanish
identically in the equilibrium state, one could easily understand
\cite{volo} why the huge condensation energy of the unperturbed
vacuum plays no role, thus obtaining an intuitive solution of the
cosmological-constant problem found in connection with the energy of
the quantum vacuum \footnote{In this sense, with this approach one
is taking seriously Feynman's indication that ``the first thing we
should understand is how to formulate gravity so that it doesn't
interact with the vacuum energy'' \cite{rule}.}.

This is not the place to discuss the various pros and cons of this
type of approach. Instead, in our context of the ether-drift
experiments, we will explore some possible phenomenological
consequence. To this end, let us assume a zeroth-order model of
gravity with a scalar field $s_0(x)$ which, at least on some
coarse-grained scale, behaves as the Newtonian potential. Then, how
could its effects be effectively re-absorbed into a curved metric
structure? At a pure kinematical level and regardless of the
detailed dynamical mechanisms, the standard basic ingredients would
be: 1) space-time dependent modifications of the physical clocks and
rods and 2) space-time dependent modifications of the velocity of
light. This point of view can be well represented by the following
two citations: \vskip 10 pt
\par\noindent Citation 1:

``It is possible, on the one hand, to postulate that the velocity of
light is a universal constant, to define {\it natural} clocks and
measuring rods as the standards by which space and time are to be
judged and then to discover from measurement that space-time is {\it
really} non-Euclidean. Alternatively, one can {\it define} space as
Euclidean and time as the same everywhere, and discover (from
exactly the same measurements) how the velocity of light and natural
clocks, rods and particle inertias {\it really} behave in the
neighborhood of large masses'' \cite{atkinson}. \vskip 10 pt
\par\noindent Citation 2:

``Is space-time really curved? Isn't it conceivable that space-time
is actually flat, but clocks and rulers with which we measure it,
and which we regard as perfect, are actually rubbery? Might not even
the most perfect of clocks slow down or speed up and the most
perfect of rulers shrink or expand, as we move them from point to
point and change their orientations? Would not such distortions of
our clocks and rulers make a truly flat space-time appear to be
curved? Yes.''\cite{thorne} \vskip 10 pt

Therefore, within this interpretation of the space-time curvature,
one might wonder about the fundamental assumption of General
Relativity that, even in the presence of gravity, the velocity of
light in vacuum $c_\gamma$ is a universal constant, namely it
remains the same, basic parameter $c$ entering Lorentz
transformations. Notice that, here, we are not considering the so
called coordinate-dependent speed of light. Rather, our interest is
focused on the value of the true, physical $c_\gamma$ as, for
instance, obtained from experimental measurements in vacuum optical
cavities placed on the earth surface.

To understand the various aspects, a good reference is Cook's
article ``Physical time and physical space in general relativity''
\cite{cook}. This article makes extremely clear which definitions of
time and length, respectively $d\tau$ and $d l$, are needed if all
observers have to measure the same, universal speed of light
(``Einstein postulate''). For a static metric, these definitions are
$d\tau^2=g_{00} dt^2$ and $dl^2=g_{ij}dx^i dx^j$. Thus, in General
Relativity, the condition $ds^2=0$, which governs the propagation of
light, can be expressed formally as \BE ds^2= c^2d\tau^2-dl^2=0 \EE
and, by construction, yields always the same universal speed
$c=dl/d\tau$.

For the same reason, however, if the physical units of time and
space were instead defined to be $d\hat \tau$ and $d\hat l$ with,
say, $d\tau = q~ d\hat\tau$ and $dl=p~ d\hat l$, the same condition
\BE ds^2= c^2q^2d \hat \tau^2-p^2 d\hat l^2=0 \EE would now be
interpreted in terms of the different speed \BE c_\gamma={{ d\hat
l}\over{ d\hat \tau }}=c ~{{q}\over{p}} \equiv {{c}\over { {\cal
N}_v }} \EE The possibility of different standards for space-time
measurements is thus a simple motivation for an effective vacuum
refractive index ${\cal N}_v\neq 1$. As we are going to illustrate,
this scenario can be tested and shown to be consistent with present
ether-drift experiments.

For sake of clarity, we shall start our analysis from the
unambiguous point of view of special relativity: the right
space-time units are those for which the speed of light in the
vacuum $c_\gamma$, when measured in an inertial frame, coincides
with the basic parameter $c$ entering Lorentz transformations.
However, inertial frames are just an idealization. Therefore the
appropriate realization is to assume {\it local} standards of
distance and time such that the identification $c_\gamma=c$ holds as
an asymptotic relation in the physical conditions which are as close
as possible to an inertial frame, i.e. {\it in a freely falling
frame} (at least by restricting light propagation to a space-time
region small enough that tidal effects of the external gravitational
potential $U_{\rm ext}(x)$ can be ignored). This is essential to
obtain an operational definition of the otherwise {\it unknown}
parameter $c$.

With this premise, as already discussed in ref.\cite{gerg}, light
propagation for an observer $S'$ sitting on the earth surface can be
described with increasing degrees of accuracy starting from step i),
through ii) and finally arriving to iii):

\begin{figure}
\begin{center}
\includegraphics[scale=0.35]{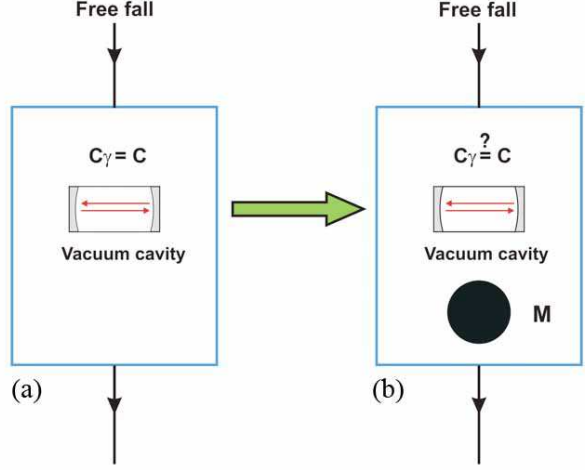}
\end{center}
\caption{ {\it A pictorial representation of the effect of a heavy
mass $M$ carried on board of a freely-falling system, case (b). With
respect to the ideal case (a), the mass $M$ modifies the local
space-time units and could introduce a vacuum refractivity so that
now $c_\gamma \neq c$.}
 } \label{freefall}
\end{figure}

~~~i) $S'$ is considered a freely falling frame. This amounts to
assume $c_\gamma=c$ so that, given two events which, in terms of the
local space-time units of $S'$, differ by $(dx, dy, dz, dt)$, light
propagation is described by the condition (ff='free-fall')
\begin{equation} \label{zero1} (ds^2)_{\rm ff}=c^2dt^2- (dx^2+dy^2+dz^2)=0~\end{equation}
 \vskip 10 pt ~~~ii) To a
closer look, however, an observer  $S'$ placed on the earth surface
can only be considered a freely-falling frame up to the presence of
the earth gravitational field. Its inclusion can be estimated by
considering $S'$ as a freely-falling frame, in the same external
gravitational field described by $U_{\rm ext}(x)$, that however is
also carrying on board a heavy object of mass $M$ (the earth mass
itself) which affects the local space-time structure, see
Fig.\ref{freefall}. To derive the required correction, let us denote
by $\delta U$ the extra Newtonian potential produced by the heavy
mass $M$ at the experimental set up where one wants to describe
light propagation. Let us also denote by ($dx$, $dy$, $dz$, $dt$)
the coordinate differences of the two chosen events (which for $M=0$
coincide with the local space-time units of the freely-falling
observer). According to General Relativity, and to first order in
$\delta U$, light propagation for the $S'$ observer is now described
by
\begin{eqnarray} \label{gr}
ds^2=c^2dt^2 (1-2{{|\delta U|}\over{c^2}})
-(dx^2+dy^2+dz^2)(1+2{{|\delta U|}\over{c^2}}) \equiv c^2 d\tau^2 -
dl^2=0
\end{eqnarray}
where $d\tau^2=(1-2{{|\delta U|}\over{c^2}})dt^2 $ and
$dl^2=(1+2{{|\delta U|}\over{c^2}})(dx^2+dy^2+dz^2) $ are the
physical units of General Relativity in terms of which one obtains
the same universal value $dl/d\tau=c_\gamma=c$.

Though, to check experimentally the assumed identity $c_\gamma =c$
one should compare with a theoretical prediction for $(c-c_\gamma)$
and thus {\it necessarily} modify some formal ingredient of General
Relativity. As a definite example, let us maintain the same
definition of the unit of length but change the unit of time. The
reason derives from the observation that physical units of time
scale as inverse frequencies and that the measured frequencies $\hat
\omega$ for $ \delta U \neq 0$, when compared to the corresponding
value $\omega$ for $\delta U = 0$, are red-shifted according to
\begin{equation} \hat \omega= (1-{{|\delta U|}\over{c^2}})~ \omega
\end{equation}
Therefore, rather than the {\it natural} unit of time
$d\tau=(1-{{|\delta U|}\over{c^2}})dt $ of General Relativity, one
could consider the alternative, but natural (see our Citation 1),
unit of time
\begin{equation} d\hat t=(1+{{|\delta
U|}\over{c^2}})~dt \end{equation} Then, to reproduce Eq.(\ref{gr}),
we can introduce a vacuum refractive index
\begin{equation} \label{lambda}  {\cal N}_v\sim 1+2{{|\delta U|}\over{c^2}}
\end{equation}
so that the {\it same} Eq.(\ref{gr}) takes the form ($dl^2\equiv
(d\hat x^2 + d\hat y^2 + d\hat z^2)$)
\begin{equation} \label{iso}ds^2 ={{c^2d\hat t^2}\over{{\cal
N}^2_v }}- (d\hat x^2 + d\hat y^2 + d\hat z^2)=0~\end{equation} This
gives
 $dl/d \hat t= c_\gamma= {{c}\over{{\cal N}_v}}$ and, for an observer
placed on the earth surface, a refractivity
\begin{equation} \label{refractive0} \epsilon_v= {\cal N}_v - 1 \sim
{{2G_N M}\over{c^2R}} \sim 1.4\cdot 10^{-9}\end{equation} $M$ and
$R$ being respectively the earth mass and radius.

Notice that, with this natural definition $d\hat t$,  the vacuum
refractive index associated with a Newtonian potential is the same
usually reported in the literature, at least since Eddington's 1920
book \cite{eddington}, to explain in flat space the observed
deflection of light in a gravitational field. The same expression is
also suggested by the formal analogy of Maxwell equations in General
Relativity with the electrodynamics of a macroscopic medium with
dielectric function and magnetic permeability \cite{volkov}
$\epsilon_{ik}=\mu_{ik}=\sqrt{-g}~{{(-g^{ik})}\over{g_{00} }}$.
Indeed, in our case, from the relations $g^{il}g_{lk}= \delta ^i_k $
, $(-g^{ik}) \sim \delta ^i_k ~g_{00}$ ,
$\epsilon_{ik}=\mu_{ik}=\delta ^i_k {\cal N}_v$ , we obtain
\begin{equation}   {\cal N}_v \sim \sqrt{-g} \sim \sqrt{
(1-2{{|\delta U|}\over{c^2}})(1+2{{|\delta U|}\over{c^2}})^3} \sim
1+2{{|\delta U|}\over{c^2}} \end{equation} A difference is found
with Landau's and Lifshitz' textbook \cite{landaufield} where the
vacuum refractive index entering the constitutive relations is
instead defined as ${\cal N}_v \sim {{1}\over{\sqrt{g_{00} }}}\sim
1+{{|\delta U|}\over{c^2}}$. This alternative definition \footnote{A
very complete set of references to these two possible alternatives
for the vacuum refractive index in gravitational field is given by
Broekaert \cite{broekaert}, see his footnote $^{3}$.} corresponds to
a different choice of the physical units and can also be taken into
account as a theoretical uncertainty. We emphasize that this
difference by a factor of 2 is not really essential. The main point
is that $c_\gamma$, as measured in a vacuum cavity on the earth
surface (panel {\bf (b)} in our Fig.\ref{freefall}), could differ at
a fractional level $10^{-9}$ from the ideal value $c$, as
operationally defined with the same apparatus in a true
freely-falling frame (panel {\bf (a)} in our Fig.\ref{freefall}). In
conclusion, this $c_\gamma - c$ difference can be conveniently
expressed through a vacuum refractivity of the form
\begin{equation} \label{refractive} \epsilon_v={\cal N}_v - 1 \sim
{{z}\over{2}}~ 1.4\cdot 10^{-9} \end{equation} where the factor
$z/2$ (with $z$= 1 or 2) takes into account the mentioned
theoretical uncertainty. \vskip 10 pt ~~~iii) Could one check
experimentally if ${\cal N}_v \neq$ 1 ? Today, the speed of light in
vacuum is assumed to be a fixed number with no error, namely 299 792
458 m/s. Thus if, for instance, this estimate were taken to
represent the value measured on the earth surface, in our picture
and in an ideal freely-falling frame there should be a slight
increase, namely $+{{z}\over{2}}(0.42)$ m/s with $z=$ 1 or 2. It
seems hopeless to measure unambiguously such a difference because
the uncertainty of the last precision measurements performed before
the ``exactness'' assumption had precisely this order of magnitude,
namely  $\pm 4\cdot 10^{-9}$ at the 3-sigma level or, equivalently,
$\pm 1.2$ m/s \cite{nist}.

Therefore, as pointed out in ref.\cite{gerg}, an experimental test
cannot be obtained from the value of the average isotropic speed
but, rather, from a possible {\it anisotropy} associated with a
theoretical difference between $c_\gamma$ and $c$. In fact, with a
preferred frame, and if ${\cal N}_v\neq 1$, an isotropic light
propagation as in Eq.(\ref{iso}) can only be valid for a special
state of motion of the earth laboratory. This provides the
definition of $\Sigma$ while for a non-zero relative velocity ${\bf
V}$ there are off diagonal elements $g_{0i}\neq 0$ in the effective
metric \cite{volkov}. The resulting two-way velocity would then be
given by Eq.(\ref{3intro}) with $\epsilon$ as in
Eq.(\ref{refractive}). On the basis of Eq.(\ref{bbasic2new}), and
for the typical $v\sim$ 370 km/s, we then expect a light anisotropy
${{|\Delta \bar{c}_\theta| }\over{c}}\sim ({\cal N}_v -1) (v/c)^2
\sim 10^{-15}$.  As a matter of fact, this prediction is consistent
with the presently most precise room-temperature vacuum experiment
of ref. \cite{schiller2015} and with the cryogenic vacuum experiment
of ref.\cite{cpt2013}. In particular, in the latter case this
measured $10^{-15}$ level was about 100 times larger than the
designed $O(10^{-17})$ short-term stability.

\section{Simulations of experiments with vacuum optical resonators}

Most recent ether-drift experiments measure the frequency shift
$\Delta \nu$ of two {\it rotating} optical resonators. To this end,
let us re-write Eq.(\ref{basic2}) as
\begin{equation} \label{basic2new}
    {{\Delta \nu (t)}\over{\nu_0}} = {{\Delta \bar{c}_\theta(t) } \over{c}}
    \sim
 \epsilon {{v^2(t) }\over{c^2}}\cos 2(\omega_{\rm rot}t
-\theta_0(t)) \end{equation} where $\omega_{\rm rot}$ is the
rotation frequency of the apparatus. Therefore one finds
\begin{equation} \label{basic3new} {{\Delta \nu (t)}\over{\nu_0}}
\sim 2{S}(t)\sin 2\omega_{\rm rot}t +
      2{C}(t)\cos 2\omega_{\rm rot}t \end{equation}with $C(t)$ and $S(t)$
      given in Eqs.(\ref{amplitude10}).

\begin{figure}
\begin{center}
\includegraphics[scale=0.35]{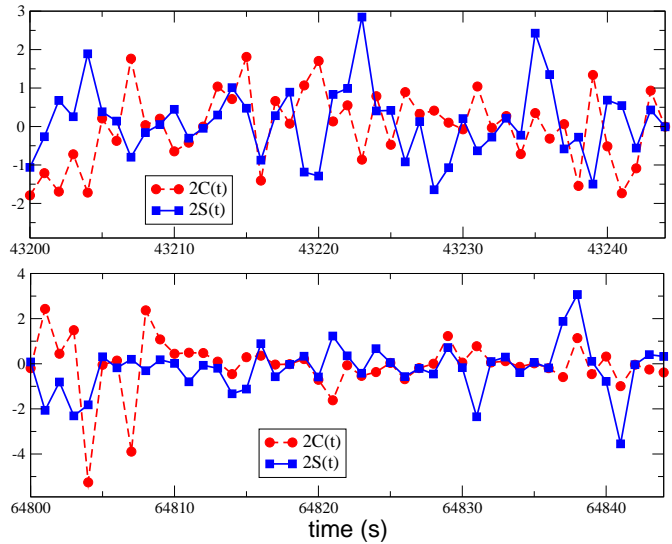}
\end{center}
\caption{\it Two typical sets of 45 seconds for the instantaneous
$2C(t)$ and $2S(t)$  in units $10^{-15}$. The two sets belong to the
same random sequence and refer to two sidereal times that differ by
6 hours. The boundaries of the stochastic velocity components in
Eqs.(\ref{vx}) and (\ref{vy}) are controlled by
$(V,\alpha,\gamma)_{\rm CMB}$ through Eqs.(\ref{projection}) and
(\ref{isot}). For a laser frequency of $2.8\cdot 10^{14}$ Hz
\cite{schiller2015}, the interval $\pm 3\cdot 10^{-15}$ corresponds
to a typical frequency shift $\Delta \nu$ in the range $\pm 1$
Hz.}\label{rotation}
\end{figure}

To estimate the signal expected with vacuum optical resonators, we
have performed several numerical simulations in the isotropic
stochastic model of Sect.3 with $\epsilon_v$ fixed as in
Eq.(\ref{refractive}) for $z=2$. However, the theoretical
uncertainty associated with the two possible choices $z=$ 1 or 2 is
also taken into account in the final formulas.

We first report in Fig.\ref{rotation} two typical sets for 2C(t) and
2S(t) during one rotation period $T_{\rm rot}=$ 45 seconds of the
apparatus of ref.\cite{newberlin}. The two sets belong to the same
random sequence and refer to two sidereal times that differ by 6
hours. The set $(V,\alpha,\gamma)_{\rm CMB}$ was adopted to control
the boundaries of the stochastic velocity components through
Eqs.(\ref{nassau1}), (\ref{projection}) and (\ref{isot}). The value
$\phi= 52$ degrees was also fixed to reproduce the average latitude
of the laboratories in Berlin and D\"usseldorf. For a laser
frequency of $2.8\cdot 10^{14}$ Hz \cite{schiller2015}, the interval
$\pm 3\cdot 10^{-15}$ of these dimensionless amplitudes corresponds
to a random instantaneous frequency shift $\Delta \nu$ in the
typical range $\pm 1$ Hz. This is well consistent with the signal
observed in ref.\cite{schiller2015}, see their Fig.4.

To compare with data extending over longer time intervals one has
first to take into account the large, long-term drift which affects
the experimental frequency shift. For instance, for the presently
most precise experiment of ref.\cite{schiller2015}, for time
variations of several hours this drift is about $\pm 500$ Hz, see
their Fig.3 (top part). This is about 1000 times larger than the
typical signal expected in our model, thus suggesting that we might
be forced to abandon altogether the possibility of a precision test
of our picture.

However, a way out derives from the observation that, although the
frequency shift changes by such a large amount, still one can
correct the data in order to achieve a much better stability.
Indeed, by suitable modeling and subtraction of the drift, the
typical variation of the shift over 1 second becomes about $\pm
0.24$ Hz (see their Fig.3, bottom part) and thus at the level $\pm 8
\cdot 10^{-16}$. This means that, after correcting the data, the
{\it local} properties of the signal, i.e. its characteristic
variations over a time scale of 1 second, depend on the possible
times $t_i$, $t_j$, $t_k$...of the observations to a negligible
extent as compared to the original differences among the
corresponding $\Delta \nu (t_i)$, $\Delta \nu (t_j)$, $\Delta \nu
(t_k)$...Then, even if these were differing by a large amount, we
can now get a test at the $10^{-15}$ level.

\begin{figure}
\begin{center}
\includegraphics[scale=0.40]{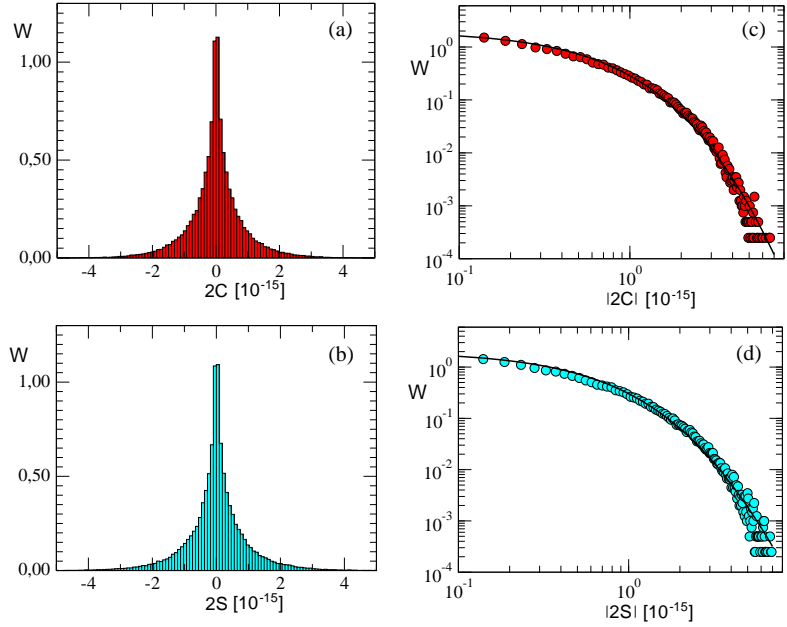}
\end{center}
\caption{ {\it We show, see (a) and (b), the histograms $W$ obtained
from a single simulation of instantaneous measurements of $\rm
2C=2C(t)$ and $\rm 2S=2S(t)$ generated at regular steps of 1 second
over an entire sidereal day. The vertical normalization is to a unit
area. The mean values are $2{\rm C}_0=\rm \langle 2C\rangle_{\rm
day} =-1.6 \cdot 10^{-18}$, $2{\rm S}_0=\rm \langle 2S\rangle_{\rm
day} =4.3 \cdot 10^{-18}$ and the standard deviations $\rm
\sigma(2C)=8.7 \cdot 10^{-16}$, $\rm \sigma(2S)=9.6\cdot 10^{-16}$.
We also show, see (c) and (d), the corresponding plots in a log-log
scale and the fits with Eq.(\ref{qexp}). The boundaries of the
stochastic velocity components in Eqs.(\ref{vx}) and (\ref{vy}) are
controlled by $(V,\alpha,\gamma)_{\rm CMB}$ through
Eqs.(\ref{projection}) and (\ref{isot}).}} \label{ourhisto}
\end{figure}

To compare with such a high short-term stability, we have thus
simulated sequences of instantaneous measurements performed at
regular steps of 1 second over an entire sidereal day. With such a
type of simulation, we can also get an idea of the $C_k$ and $S_k$,
entering Eqs.(\ref{amorse1}) and (\ref{amorse2}), for a large but
finite statistics (where one cannot get exactly zero as expected
from Eqs.(\ref{vanishing}) and (\ref{vanishing2})).  For a
particular random sequence, the resulting histograms of $2C$ and
$2S$ are reported in panels (a) and (b) of Fig.\ref{ourhisto}.

In view of Eqs.(\ref{vanishing}) and (\ref{vanishing2}) the non-zero
averages $ \langle 2C\rangle_{\rm day}=2C_0={\cal O}(10^{-18})$, $
\langle 2S\rangle_{\rm day}=2S_0={\cal O}(10^{-18})$ should only be
considered as statistical fluctuations around zero. The same holds
true for the other $C_k$ and $S_k$ Fourier coefficients in
Eqs.(\ref{amorse1}) and (\ref{amorse2}). By fitting the generated
distributions to Eqs.(\ref{amorse1}) and (\ref{amorse2}) one gets
values which are also ${\cal O}(10^{-18})$ or smaller and which
fluctuate randomly around zero as expected. This simulated pattern
is in complete agreement with the typical magnitude ${\cal
O}(10^{-18})$ obtained in ref.\cite{schiller2015} from the
experimental data.

On the other hand, in such simulations of one-day measurements at
steps of one second, the standard deviations around the $10^{-18}$
averages, say $\sigma_{\rm th}(2C)$ and $\sigma_{\rm th}(2S)$, are
very stable ($z=$ 1 or 2)
\begin{equation} \label{sigmas}
\left[ \sigma_{\rm th}(2C) \right]_{\rm 1-second } \sim
{{z}\over{2}}(8.7 \pm 0.8)\cdot 10^{-16}
 ~~~~~~~~~~~~~~~~ \left[ \sigma_{\rm th}(2S) \right]_{\rm 1-second }
 \sim {{z}\over{2}}(9.6 \pm 0.9)\cdot 10^{-16}
\end{equation} Here the $\pm$ uncertainties reflect the observed
variations due to the truncation of the Fourier modes in
Eqs.(\ref{vx}), (\ref{vy}) and to the dependence on the random
sequence.

From Eq.(\ref{basic3new}), by combining quadratically these two
sigma's, we estimate
\begin{equation}
\left[\sigma_{\rm th}({{\Delta \nu }\over{\nu_0}})\right]_{\rm
1-second }\sim {{z}\over{2}} (9\pm 1) \cdot 10^{-16}\end{equation}
so that, for a laser frequency $\nu_0=2.8\cdot 10^{14}$ Hz
\cite{schiller2015}, we expect a typical spread
\begin{equation}
\left[ \sigma_{\rm th}(\Delta \nu) \right]_{\rm 1-second } \sim
{{z}\over{2}}(0.26 \pm 0.02) ~{\rm Hz}\end{equation} of the
frequency shift measured every 1 second over a one-day period. For
$z=2$, this estimate is in very good agreement with the mentioned
experimental value
\begin{equation} \left[ \sigma_{\rm exp}(\Delta \nu) \right]_{\rm
1-second }\sim 0.24~ {\rm Hz}
\end{equation} which is reported in ref.\cite{schiller2015} for an
integration time of 1 second. Therefore, to the present best level
of accuracy, this agreement strongly favours the value $z=2$, which
is the only free parameter of our scheme.

Our estimates are also well consistent with the analogous (but
slightly less stringent) limit $\sigma_{\rm exp}({{\Delta \nu
}\over{\nu_0}})\sim 1.5\cdot 10^{-15}$, at $1\div 2$ seconds, placed
by the cryogenic experiment of ref.\cite{cpt2013}. Notice that this
measured $10^{-15}$ level was about 100 times larger than the
expected ${\cal O}(10^{-17})$ short-term stability. However, by the
authors of ref.\cite{cpt2013}, it was interpreted as a spurious
effect due to a lack of rigidity of their cryostat. Probably, they
have not considered the possibility of a genuine random signal and
of intrinsic limitations placed by the vacuum structure.

We emphasize that the generated distributions are very different
from a Gaussian shape, an aspect which is characteristic of
probability distributions for instantaneous data in turbulent flows
(see e.g. \cite{sreenivasan,beck}). To better appreciate the
deviation from Gaussian behavior, in panels (c) and (d) we plot the
same data in a log$-$log scale. The resulting distributions are well
fitted by the so-called $q-$exponential function \cite{tsallis}
\begin{equation}
\label{qexp} f_q(x) = a (1 - (1 - q) x b)^{1 / (1 -
q)}\end{equation}  with ``entropic'' index $q\sim 1.1$. This
explains why, by performing extensive simulations, there might be
occasionally large spikes of the instantaneous amplitude, up to $ 7
\cdot 10^{-15}$ or larger, when many Fourier modes sum up coherently
(see the tails in panels (c) and (d) of Fig.\ref{ourhisto}). The
effect of these spikes, which lie at about 7 sigma's in terms of the
standard deviations Eq.(\ref{sigmas}), gets smoothed when averaging
but their non-negligible presence (about $10^{-4}$ probability) is
characteristic of the stochastic model. Otherwise, for a Gaussian
distribution, 7 sigma's would correspond to a $10^{-11}$
probability.


As already observed for the classical experiments, another reliable
indicator is the statistical average of the quadratic amplitude of
the signal \[ A(t)\equiv 2\sqrt{S^2(t) + C^2(t)} \] which is a
positive-definite quantity and, as such, remains definitely non-zero
after any averaging procedure. In this case, by using Eqs.
(\ref{refractive}) and (\ref{vanishing}), one finds ($z=$ 1 or 2)
\begin{equation} \label{reduction}\langle A^{\rm th}(t) \rangle_{\rm
stat}=\epsilon_v
 {{\tilde{v}^2(t)}\over{c^2}}
~{{1}\over{3}} \sum^{\infty}_{n=1} {{1}\over{n^2}}={{z}\over{2}}~
(7.7 \cdot 10^{-16}) {{\tilde{v}^2(t)}\over{(300~{\rm km/s})^2}}
\end{equation}

By maintaining the CMB parameters $(V,\alpha,\gamma)_{\rm CMB}$ and
fixing $\phi= 52$ degrees, one gets a daily average $\sqrt {\langle
\tilde {v}^2 \rangle}_{\rm day} \sim 332$ km/s from the relation
\cite{gerg} \begin{equation} \label{daily} \langle \tilde{v}^2
\rangle_{\rm day}= V^2
       \left(1- \sin^2\gamma\sin^2\phi
       - {{1}\over{2}} \cos^2\gamma\cos^2\phi \right) \end{equation}
Thus, we predict a daily average amplitude ($z=$ 1 or 2)
\begin{equation} \label{amplitudedaily} \langle A^{\rm th} \rangle_{\rm
day}\sim {{z}\over{2}}~9\cdot 10^{-16} \end{equation} that, for a
laser frequency $2.8\cdot 10^{14}$ Hz, corresponds again to a
typical instantaneous frequency shift $|\Delta \nu|_{\rm th} \sim
{{z}\over{2}} $ 0.26 Hz.

Other tests of the model will be possible if, besides the results of
fits to the standard parameterizations Eqs. (\ref{amorse1})
and(\ref{amorse2}), also the basic instantaneous amplitudes $A(t)$,
$S(t)$ and $C(t)$ will become available. By comparing with these
genuine data, we could also get other insights and improve on our
simplest model of stochastic turbulence.

To conclude, we observe that a crucial test of our model consists in
detecting tiny daily variations of the amplitude. This is a very
difficult task due to the necessity of subtracting the mentioned
systematic long-term drift which is much larger than the variation
of a small fraction of Hz expected in our picture. Nevertheless,
assuming that this subtraction could be done unambiguously to
appreciate differences at the relative level $10^{-16}$, for the CMB
parameters at the latitude of Berlin-D\"usseldorf, where the scalar
velocity $\tilde{v}(t)$ in Eq.(\ref{projection}) changes in the
range $260\div 370$ km/s, from Eq.(\ref{reduction}) we expect the
typical range ($z=$ 1 or 2)
\begin{equation} \langle A^{\rm th}(t) \rangle_{\rm stat}={{z}\over{2}}~(9
\pm 3)\cdot 10^{-16}
\end{equation} More generally, if a daily variation of the amplitude
will be detected, one could try to fit from the data the kinematical
parameters $(V,\alpha,\gamma)$ entering Eq.(\ref{projection}).

\section{Gaseous media vs. vacuum and solid dielectrics}

\subsection{Light anisotropy in gases as a non local thermal effect}

Now, returning to the gas case, it is natural to ask: independently
of all symmetry arguments, why there should be a non-zero light
anisotropy in the earth laboratory where (the container of) the gas
is at rest? Moreover, only the region of refractive index
infinitesimally close to the ideal vacuum ${\cal N}=1$ has been
analyzed. What about experiments performed in the other region where
${\cal N}$ differs substantially from unity, as in solid
dielectrics?

By concentrating on the first question, our explanation will consist
in a non-local temperature gradient of a fraction of millikelvin
associated with the earth motion. This idea comes out naturally by
recalling that from the relation \BE \left[
{{\Delta\bar{c}_\theta}\over{c}} \right]_{\rm gas} \sim ( {\cal
N}_{\rm gas}-1)~{{v^2}\over{c^2}} \cos 2\theta \EE and correcting
with the different refractive indexes, respectively $({\cal N}_{\rm
air}-1)\sim 2.8\cdot 10^{-4}$ and $({\cal N}_{\rm helium}-1) \sim
3.3\cdot 10^{-5}$, the same typical {\it kinematical} velocity
$v\sim$ 300 km/s can account for the observed light anisotropy,
namely ${{|\Delta\bar{c}_\theta|}\over{c}} ={\cal O}(10^{-10})$ for
air and to ${{|\Delta\bar{c}_\theta|}\over{c}} ={\cal O}(10^{-11})$
for helium. Therefore since, for all practical purposes, a possible
non-zero vacuum anisotropy $({\cal N}_v-1)(v/c)^2\lesssim 10^{-15}$
is irrelevant, the answer to our first question requires to find the
mechanism which {\it enhances} substantially the anisotropy in the
gas case.

To this end, it is natural to exploit the traditional {\it thermal}
interpretation of the residuals of the classical experiments. This
old argument, which gave the main motivation for Kennedy's
replacement of air with gaseous helium in his optical paths, will
now be illustrated by the explicit calculation of the temperature
dependence of the gas refractive index ${\cal N}$.

The starting point is the Lorentz-Lorentz equation (see e.g.
\cite{stone})
\begin{eqnarray}\label{refra1}
        {{ {\cal N}^2 -1}\over{ {\cal N}^2 + 3  }} = A_R \rho + B_R
        \rho^2 ...
\end{eqnarray}
where $\rho$ is the molar density and $A_R= (4/3) \pi N_A \alpha$ is
expressed in terms of the Avogadro number $N_A$ and of the molecular
polarizability $\alpha$. The coefficient $B_R$ takes into account
two-body interactions and in our case, of air and helium at
atmospheric pressure, this higher order term is completely
negligible. Since ${\cal N}$ is very close to unity, we obtain the
simplified formula for the gas refractivity
\begin{eqnarray}\label{refra2}
       \epsilon= {\cal N} -1 \sim  {{ 3}\over{ 2 }}A_R \rho
\end{eqnarray}
In the ideal-gas approximation, the molar density at Standard
Temperature and Pressure (atmospheric pressure and zero centigrade
or 273.15 K) has the well known value
\begin{eqnarray}\label{refra3}
        \rho(STP)=  {{ P}\over{ RT }} ={{ 101325}\over{ (8.314)
        (273.15)}}~{ \rm mol }\cdot { \rm m}^{-3 }\sim 4.46 \cdot 10^{-5 }~{ \rm mol }\cdot{ \rm cm}^{-3 }
\end{eqnarray}
Thus, for instance, for helium at STP and a wavelength $\lambda=$
633 nm, where $A_R\sim 0.52~ { \rm mol }^{-1 }\cdot{ \rm cm}^{3 }$
\cite{stone}, one finds $\epsilon\sim 3.5\cdot 10^{-5 }$.

The interesting aspect is that, in the ideal-gas approximation, the
variation of the refractivity with the temperature has the very
simple expression
\begin{eqnarray}\label{refra4}
-{{ \partial \epsilon }\over{ \partial T }} \sim {{ 3}\over{ 2 }}
A_R {{ P}\over{ RT^2 }} \sim {{ \epsilon} \over{ T }}
\end{eqnarray}
\begin{table*}
\caption {\it The average 2nd-harmonic amplitude observed in various
classical ether-drift experiments and the resulting temperature
difference obtained from  Eqs.(\ref{newintroair}) and
(\ref{newintrohelium}).}
\begin{center}
\begin{tabular}{cllll}
\hline Experiment &gas
&~~~~$A^{\rm EXP}_2$ &~~~ ${{2D}\over{\lambda}}$~~~~&  $\Delta T^{\rm EXP}$({\rm mK})\\
\hline
Michelson-Morley(1887)   & air & $(1.6 \pm 0.6)\cdot 10^{-2 }$&~~~$4\cdot 10^7$ &$0.40 \pm 0.15$\\
Miller(1925-1926)  & air& $ (4.4 \pm 2.2)\cdot 10^{-2 } $ & ~~ $ 1.12\cdot 10^8$&$0.39 \pm 0.20$ \\
Illingworth(1927) & helium & $ (2.2 \pm 1.7)\cdot 10^{-4}  $  &~~~$7 \cdot 10^6$& $0.29\pm 0.22$\\
Tomaschek (1924) & air & $(1.0\pm 0.6)\cdot 10^{-2 }  $ &~~~$3\cdot 10^7$& $0.33 \pm 0.20$\\
Piccard-Stahel(1928)&air &$(2.8 \pm 1.5)\cdot 10^{-3 }$  &  ~~~$ 1.28 \cdot 10^7$& $0.22 \pm 0.12$\\
Joos(1930)  &helium&$(1.4 \pm 0.8)\cdot 10^{-3 }   $  & ~~ $7.5 \cdot 10^7$&$0.17 \pm 0.10$\\
\hline
\end{tabular}
\end{center}
\end{table*}

Therefore, by recalling the definition Eq.(\ref{refractivetheta}), a
small temperature difference $\Delta T(\theta)$ induces a light
anisotropy of typical magnitude \BE  {{|\Delta \bar{c}_\theta| }
\over{c}} \sim ~ |\bar{\cal N}(\theta) -\bar{\cal N}(\pi/2+ \theta)|
\sim {{~ \epsilon |\Delta T(\theta)|}  \over{ T }}\EE We can thus
extract an experimental temperature difference from the 2nd-harmonic
amplitudes $A_2$ in the fringe shifts
\begin{equation}
\label{newintroacca} {{\Delta \lambda(\theta)}\over{\lambda}} \sim
{{2D}\over{\lambda}} ~{{\Delta \bar{c}_\theta } \over{c}}= A_2 \cos
2 \theta
\end{equation}
At room temperature, say $T= 288$ K$ ~+\Delta T$, this gives the
relations
\begin{equation}
\label{newintroair} A^{ \rm EXP}_2( { \rm air} ) \sim
{{2D}\over{\lambda}}~ {{\epsilon_{\rm air}(T) \Delta T ^{\rm
EXP}}\over{T}} \sim {{2D}\over{\lambda}} \cdot 10^{-9 } ~{{ \Delta T
^{\rm EXP}} \over{ {\rm mK} }}
\end{equation} and
\begin{equation}
\label{newintrohelium} A^{\rm EXP}_2({\rm helium})
\sim{{2D}\over{\lambda}} ~{{\epsilon_{\rm helium}(T) \Delta T ^{\rm
EXP}}\over{T}} \sim  {{2D}\over{\lambda}} ~( 1.1 \cdot 10^{-10
})~{{\Delta T^{\rm EXP}} \over{ {\rm mK}  }}
\end{equation}
The temperature differences from the various experiments are
reported in Table 3.

Our calculation shows that the old estimates of $1\div 2$ mK by
Kennedy, Shankland and Joos (see \cite{shankland}) were too large,
by about one order of magnitude. At the same time, the six
determinations in Table 3 are well consistent with each other as
shown by the excellent chi-square, $ 2.4/(6-1)=0.48$, of their
average \BE \langle \Delta T^{\rm EXP} \rangle = (0.26 \pm 0.06) ~
{\rm mK} \EE Thus the light anisotropy observed in the classical
experiments could also be explained in terms of a thermal gradient
with a {\it non-local} origin suggesting that $\langle \Delta T^{\rm
EXP} \rangle$ might ultimately be related to the CMB temperature
dipole of $\pm 3$ mK. However, without a quantitative calculation,
the fundamental energy flow expected in a Lorentz-non-invariant
vacuum (see the Appendix) could represent another possible
explanation. In any case, a thermal interpretation had already been
deduced in refs.\cite{plus,foop} by noticing that
Eq.(\ref{bbasic2new}) is just a special case of the light anisotropy
expected from convective currents of the gas molecules associated
with the earth motion (see Appendix 1 of \cite{foop}). All this, can
most easily be summarized by re-expressing the $\theta-$dependent
refractive index ${\bar{\cal N}(\theta)}\equiv {{c}\over{\bar
c_\gamma (\theta)}}$ Eq.(\ref{refractivetheta}) as
\begin{eqnarray}\label{6intro}
       {{ \bar{\cal N}_{\rm gas}(\theta)}\over { {\cal N}_{\rm gas}}} \sim 1+(a_{\rm thermal}+ a_{ v})
        \beta^2\left(2 -
       \sin^2\theta\right)
\end{eqnarray}
Here $a_{\rm thermal}\equiv ({\cal N}_{\rm gas}- {\cal N}_v )$ and
$a_{ v}\equiv ({\cal N}_v-1)\sim 10^{-9}$. In this way, for gases in
normal conditions, the genuine $a_v$ vacuum term is always
numerically irrelevant and all anisotropy is due to $a_{\rm
thermal}$ (recall that $a_{\rm thermal}\sim 2.8 \cdot 10^{-4}$ or
$a_{\rm thermal}\sim 3.3 \cdot 10^{-5}$ respectively for air or
gaseous helium at room temperature and atmospheric pressure).

\subsection{Light anisotropy in solid dielectrics}

Armed with this type of thermal interpretation, we can then address
the second question concerning the ether-drift experiments in solid
dielectrics, of the type performed by Shamir and Fox \cite{fox} in
1969. They were aware that the Michelson-Morley experiment did not
yield a strictly zero result: ``The non-zero result might have been
real and due to the fact that the experiment was performed in air
and not in vacuum'' \cite{fox}. Thus, with ${\cal N}$ values
substantially above unity, and within the traditional
Lorentz-contraction interpretation, one might expect to observe a
large ether-drift ${{|\Delta\bar{c}_\theta|}\over{c}}\sim ({\cal
N}^2-1)\beta^2\sim \beta^2\sim 10^{-6}$. The search for such effect
was the motivation for their experiment in perspex (${\cal N}=1.5$).
Since no such enhancement was observed, they concluded that the
experimental basis of special relativity was strengthened.

However, with a thermal interpretation of the residuals observed in
gaseous media, the two different behaviors can be reconciled. In a
strongly bound system as a solid, in fact, a small temperature
gradient of a fraction of millikelvin would mainly dissipate by heat
conduction without generating any appreciable particle motion or
light anisotropy in the rest frame of the apparatus. Hence, the
non-trivial, physical difference between experiments in gaseous
systems and experiments in solid dielectrics. In the latter case, we
do not expect any enhancement with respect to the pure vacuum case.
This means that, with very precise measurements, a fundamental
vacuum anisotropy  $\lesssim 10^{-15}$ should also show up in
strongly bound solid dielectrics.

To see this, let us first recall that, in our picture, a relation
similar to Eq.(\ref{6intro}) is also valid in the vacuum limit where
it takes the form
\begin{eqnarray}
\label{5intro}
        {{ \bar {\cal N}_v(\theta)}\over {{\cal N}_v}} \sim 1+ ({\cal N}_v-1) \beta^2\left(2 -
       \sin^2\theta\right)
\end{eqnarray}
with, see Eq.(\ref{refractive}), $({\cal N}_v-1)\sim
{{z}\over{2}}~1.4 \cdot 10^{-9}$ and $z=$ 1 or 2.

The existence of ${\cal N}_v$  produces a tiny difference between
the refractive index defined relatively to the ideal value
$c_\gamma=c$ and the refractive index defined relatively to the
physical isotropic value $c_\gamma=c/{\cal N}_v$ measured on the
earth surface. The percentage difference between the two definitions
is proportional to ${\cal N}_v-1$ and, for all practical purposes,
can be ignored.

More significantly, all materials exhibit a background vacuum
anisotropy proportional to $({\cal N}_v-1)\beta^2\sim 10^{-15}$. As
explained, for gases in normal pressure conditions this genuine
vacuum effect can be neglected. For solid dielectrics, on the other
hand, where no thermal enhancement is expected, one should keep
track of the vacuum term. To this end, first replace the average
isotropic value \BE {{c}\over { {\cal N}_{\rm solid}}} \to {{c}\over
{ {\cal N}_v {\cal N}_{\rm solid} }}\EE Then use Eq.(\ref{5intro})
to replace ${\cal N}_v$ in the denominator with $\bar {\cal
N}_v(\theta)$ and take into account the motion of the laboratory
relatively to the preferred $\Sigma$ frame. This is equivalent to
define a $\theta-$dependent refractive index for the solid
dielectric
\begin{eqnarray}\label{7intro}
        {{  \bar{\cal N}_{\rm solid}(\theta)}\over { {\cal N}_{\rm solid}}} \sim  1+({\cal N}_v-1) \beta^2\left(2 -
       \sin^2\theta\right)
\end{eqnarray}
so that
\begin{equation}
\label{refractivetheta1} \left[ {\bar c_\gamma (\theta)}
\right]_{\rm solid}={{c}\over{\bar{\cal N}_{\rm
solid}(\theta)}}={{c}\over { {\cal N}_{\rm solid}}} \left[ 1-({\cal
N}_v-1) \beta^2\left(2 -
       \sin^2\theta\right)\right]
\end{equation}
with an anisotropy
\begin{equation}
{{ \left[\Delta\bar{c}_\theta\right]_{\rm solid}} \over {\left[ c/
{\cal N}_{\rm solid}\right] }} \sim({\cal N}_v-1) \beta^2
\cos2\theta \sim{{z}\over{2}}~1.4 \cdot 10^{-9}\cdot 10^{-6}
\cos2\theta \lesssim 10^{-15}
\end{equation}
Thus, for light propagation in solids, we would predict the same
type of irregular signal discussed for pure vacuum and shown in our
Fig.\ref{rotation}. This expectation is consistent with the other
cryogenic experiment by Nagel et al. \cite{nagelnature}. In fact,
most electromagnetic energy propagates in a dielectric with
refractive index ${\cal N}\sim 3$ (at microwave frequencies) but the
typical, {\it instantaneous} determination (see their Fig.3 b) is
again $|{{\Delta \nu }\over{\nu_0}}|\lesssim 10^{-15}$ as in the
vacuum case (and as in the vacuum case it is about 1000 times larger
than the average determination $|\langle{{\Delta \nu }\over{\nu_0}}
\rangle| \lesssim 10^{-18}$ obtained by combining a large number of
measurements). For this reason the persistence, in vacuum and in
solid dielectrics, of the irregular $10^{-15}$ signal should be
definitely established. Instead, the present statistical averages,
at the level $|\langle{{\Delta \nu }\over{\nu_0}} \rangle |\lesssim
10^{-18}$, have no particular significance. With a stochastic
signal, there is no problem in reaching the level $|\langle{{\Delta
\nu }\over{\nu_0}} \rangle| \lesssim 10^{-19}$, $|\langle{{\Delta
\nu }\over{\nu_0}} \rangle |\lesssim 10^{-20}$ ... by simply
increasing the number of observations.

Finally, a complementary test could be performed by placing the
vacuum (or solid dielectric) optical cavities on board of a
satellite, as in the OPTIS proposal \cite{optis}. In this case
where, even in a flat-space picture, the effective vacuum refractive
index ${\cal N}_v$ for the freely-falling observer is exactly unity,
the typical instantaneous frequency shift should be much smaller (by
orders of magnitude) than the corresponding $10^{-15}$ value
measured with the same interferometer on the earth surface.

\section{Summary and conclusions}

The standard interpretation of the dominant CMB dipole anisotropy is
in terms of a Doppler effect due to the motion of the solar system
with an average velocity of 370 km/s toward a point in the sky of
right ascension 168 degrees and declination -7 degrees. As discussed
in the Introduction, the implications of this result may be more
radical than usually believed. Indeed, the satisfactory kinematic
reconstruction of the observed dipole, from the various peculiar
motions which are involved, leads to the natural concept of a global
frame of rest determined by the average distribution of matter in
the universe. At the same time, this global frame could also reflect
a vacuum structure with some degree of substantiality and, in this
sense, could characterize non-trivially the form of relativity which
is physically realized in nature.

Starting from these premises, it is natural to explore the
possibility to correlate ether-drift measurements in laboratory with
direct CMB observations in space. The present view is that no such
meaningful correlation has ever been observed. In fact all data
(from Michelson-Morley until the modern experiments with optical
resonators) are considered as a long sequence of null results
obtained in measurements with better and better systematics.

Instead, we have argued that this present view is far from obvious.
The main argument is based on a modern version of Maxwell's original
calculation for the anisotropy of the two-way velocity of light. By
using simple symmetry arguments, in the infinitesimal region of
refractive index ${\cal N}= 1 + \epsilon$ a possible non-zero
anisotropy should scale as ${{|\Delta\bar{c}_\theta|}\over{c}}\sim
\epsilon v^2/c^2$, see Eq.(\ref{bbasic2new}), $v$ being the earth
velocity with respect to a hypothetical preferred frame. Therefore,
due to the strong suppression, with respect to the classical
estimate ${{|\Delta\bar{c}_\theta|}\over{c}}\sim v^2/c^2$, the size
of the small residuals observed in the classical experiments in
gaseous media (Michelson-Morley, Miller, Illingworth, Joos,..) can
become consistent with the typical value $v\sim$ 370 km/s obtained
from the direct CMB observations. The essential point is contained
in the relation $ v^2_{\rm obs} \sim 2\epsilon v^2$ which connects
the {\it kinematical} velocity $v$ to a much smaller {\it
observable} velocity $v_{\rm obs}$ which determines the magnitude of
the fringe shifts.

For the full consistency of this interpretation, however, a change
of perspective is needed. Namely, the irregular character of the
data requires that the local velocity field $v_\mu$ which determines
light anisotropy, and as such the fringe shifts in the old
experiments or the frequency shifts in the modern experiments,
should {\it not} be identified with the global velocity field
$\tilde v_\mu$ as directly fixed by the earth cosmic  motion.
Instead, from general arguments related to the idea of the vacuum as
an underlying stochastic medium, we have proposed that the relation
between these two quantities might be indirect and similar to what
happens in turbulent flows. This means that the local $v_\mu$ could
fluctuate randomly while the global $\tilde v_\mu$ would just fix
its typical boundaries. Thus, if turbulence becomes homogeneous and
isotropic at small scales, one has a definite model where a genuine
instantaneous signal can well coexist with vanishing statistical
averages for all vectorial quantities.

In this alternative picture, the direction of the local drift in the
plane of the interferometer is a completely random quantity which
has no definite limit by combining a large number of observations.
Therefore, one should concentrate on the positive-definite quadratic
amplitude of the signal and on its time modulations. In this case,
by restricting to the amplitude, the results of ref.\cite{plus}
indicate a good consistency of the residuals of the classical
experiments with the direct CMB observations. This alternative view
should be checked with a new series of tests in which the optical
resonators, which are coupled to the lasers, are filled by gaseous
media. This would reproduce the physical conditions of those early
measurements with today's greater accuracy. At present, a first
rough check can be obtained from the time variations of a few kHz
observed in the only modern experiment performed in similar
conditions, namely the 1963 MIT experiment by Jaseja et. al
\cite{jaseja} with He-Ne lasers (see the discussion given in
\cite{epl}).

Waiting for these new experiments, we have compared our picture with
the frequency shift detected in modern vacuum experiments. The point
is that for the physical vacuum the ideal equality ${\cal N}_v =1$
might not be exact. For instance, as proposed in \cite{gerg},  a
very tiny value ${\cal N}_v - 1= \epsilon_v \sim 10^{-9}$ could
reveal the different effective refractivity between an apparatus in
an ideal freely-falling frame and an apparatus on the earth surface.
In this case, we would expect a genuine, stochastic frequency shift
${{|\Delta \nu(t)| }\over{\nu_0}} \sim \epsilon_v (v/c)^2 \sim
10^{-15}$ which coexists with vanishing statistical averages for all
vectorial quantities, such as the $C_k$ and $S_k$ Fourier
coefficients extracted from a standard temporal fit to the data with
Eqs.(\ref{amorse1}) and (\ref{amorse2}).

The numerical simulations shown in Sect.6 indicate that this
expectation is well consistent with the presently most precise
room-temperature experiment of ref.\cite{schiller2015} and with the
cryogenic experiment of ref.\cite{cpt2013} (which is only less
precise by about a factor of 2). By itself, this substantial
agreement between experiments with different systematics indicates
that the observed signal might have a genuine physical component and
not just originate from spurious noise in the spacers and the
mirrors of the optical resonators, as assumed so far. In fact, the
estimates of these contributions \cite{numata} are based on the
fluctuation-dissipation theorem and thus there is no obvious reason
that experiments operating at so different temperatures exhibit the
same instrumental effects. The unexplained agreement with
ref.\cite{cpt2013} is particularly striking in view of the factor
100 which exists between observed signal $10^{-15}$ and designed
short-term stability ${\cal O}( 10^{-17})$. Tentatively, the authors
of \cite{cpt2013} interpreted this discrepancy as being due to a
lack of rigidity of their cryostat but, probably, they have not
considered the possibility of a genuine random signal and of
intrinsic limitations placed by the vacuum structure. In this
different perspective, the alternative interpretation proposed in
\cite{gerg}, and implemented here, should also be taken into
account.

This becomes even more true in view of the very good agreement
obtained between the experimental value for the spread of the
instantaneous signal found in ref.\cite{schiller2015}, namely
$\sigma_{\rm exp}(\Delta \nu)\sim$ 0.24 Hz, and our corresponding
simulated value $\sigma_{\rm th}(\Delta \nu)\sim (0.26\pm 0.02)$ Hz
for that experiment, with $z=2$ in Eq. (\ref{refractive}).

\begin{figure}
\begin{center}
\includegraphics[scale=0.4]{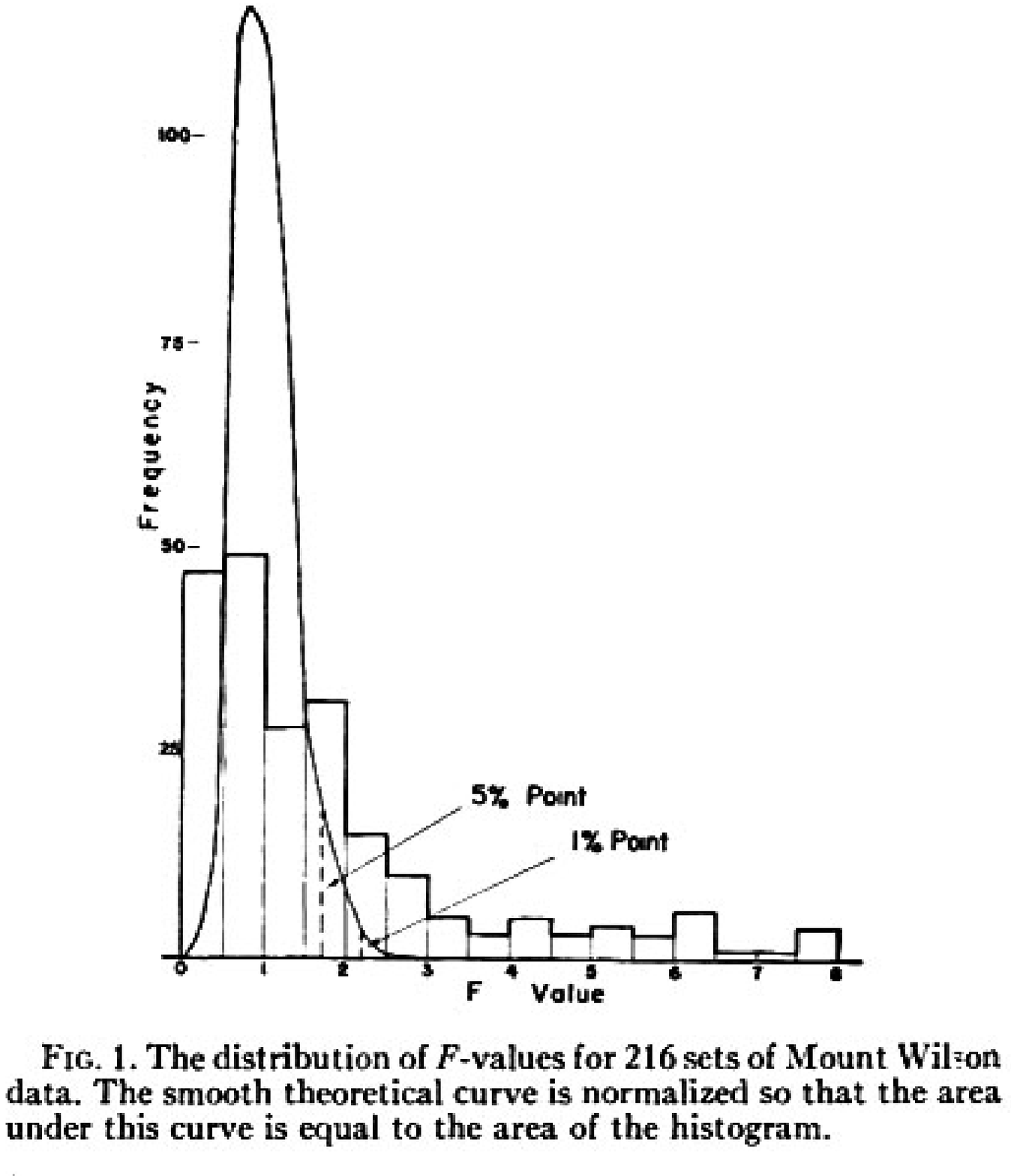}
\end{center}
\caption{\it {The probability histogram for 216 sets of Miller's
observations as computed by Shankland et al. \cite{shankland}.} }
\label{shankland}
\end{figure}

The agreement we have obtained looks very promising and opens the
possibility to reconstruct the CMB dipole with precise optical
measurements performed within the earth laboratory and thus
definitely clarify the fundamental issue of a preferred frame. To
this end, however, real {\it data} (and not just the results of {\it
fits}) should become available. In fact, our model, besides implying
vanishing statistical averages for all vectorial quantities, in
agreement with the observations, makes other definite predictions.
For instance, precise time modulations of the quadratic amplitude of
the signal and non-Gaussian (i.e. long-tail) distributions for the
individual measurements. Although, at present, modern experiments
give no information on these aspects, this idea of long tails finds
definite support in the statistical analysis of Miller's extensive
observations, see Fig.1 of the paper by Shankland et al.
\cite{shankland} reported here as our Fig.\ref{shankland}.

Finally, in Sect.7, we have addressed the possible physical
mechanism which enhances the signal in gaseous media, respectively
${{|\Delta\bar{c}_\theta|}\over{c}}= {\cal O}(10^{-10})$ and
${{|\Delta\bar{c}_\theta|}\over{c}}= {\cal O}(10^{-11})$ for air or
helium at atmospheric pressure, relatively to the instantaneous
vacuum value ${{|\Delta\bar{c}_\theta|}\over{c}}\lesssim 10^{-15}$
found in modern experiments on the earth surface. For instance, one
could imagine a suitable interaction of the incoming radiation with
the medium to produce a different polarization in different
directions. Any such mechanism, however, should act in both gaseous
matter and solid dielectrics with the final result that light
anisotropy should always increase with the refractivity of the
medium, in contrast with the experimental evidence.

Therefore, if the enhancement observed in gases has to be specific
of such weakly bound forms of matter, the natural interpretation is
in terms of a {\it non local} temperature gradient associated with
the earth motion. This shows up in all classical experiments, in
agreement with the traditional thermal interpretation of the
observed residuals. Only, its average magnitude $\langle \Delta T
\rangle = (0.26 \pm 0.06)$ mK is somewhat smaller than the old
estimates (about $1\div 2$ mK) by Kennedy, Joos and Shankland.
Conceivably, it might ultimately be related to the CMB temperature
dipole of $\pm 3$ mK or reflect the fundamental energy flow expected
in a Lorentz-non-invariant vacuum state. While, at present, we have
no definite quantitative insight, yet such thermal interpretation is
important to understand the differences and the analogies among
experiments in gaseous media, in vacuum and in solid dielectrics.
Indeed, for experiments with optical cavities maintained in an
extremely high vacuum (both at room temperature and in the cryogenic
regime), where any residual gaseous matter is totally negligible,
such tiny temperature variations cannot produce any observable
effect.

On the other hand, in solid dielectrics a so small temperature
gradient should mainly dissipate by heat conduction without
generating any appreciable particle motion or light anisotropy in
the rest frame of the apparatus. Hence, in solid dielectrics we do
not expect any sizeable enhancement with respect to what is observed
in the pure vacuum case.  This expectation is consistent with the
cryogenic experiment by Nagel et al. \cite{nagelnature} where light
propagates in a dielectric with refractive index ${\cal N}\sim 3$
(at microwave frequencies) but the typical, {\it instantaneous}
determination (see their Fig.3 b) is again
${{|\Delta\bar{c}_\theta|}\over{c}}\lesssim 10^{-15}$ as in the
vacuum case (and as in the vacuum case is about 1000 times larger
than the average determination
$|\langle{{\Delta\bar{c}_\theta}\over{c}} \rangle| \lesssim
10^{-18}$ obtained by combining a large number of measurements).

At present, our prediction of a fundamental irregular signal
${{|\Delta\bar{c}_\theta|}\over{c}}\lesssim 10^{-15}$ is the only
explanation for this observed agreement between so different
experiments, namely ref.\cite{schiller2015} with vacuum cavities at
room temperature, vs. ref.\cite{nagelnature} performed in a solid
dielectric in the cryogenic regime. The definite persistence of such
signal would confirm the existence of a fundamental preferred frame
for relativity and would have substantial implications for our
interpretation of non-locality in the quantum theory. Once
definitely established, complementary tests should be performed by
placing the vacuum (or solid dielectric) optical cavities on board
of a satellite, as in the OPTIS proposal \cite{optis}. In this ideal
free-fall environment, the typical instantaneous frequency shift
should be much smaller (by orders of magnitude) than the
corresponding $10^{-15}$ value measured with the same
interferometers on the earth surface.

\setcounter{equation}{0}
\renewcommand{\theequation}{A\arabic{equation}}


\vskip 100 pt \centerline{\bf{\LARGE Appendix }} \vskip 15 pt

\par\noindent According to general quantum field theoretical arguments (see
e.g. \cite{cpt}), deciding on the Lorentz invariance of the vacuum
state $|\Psi^{(0)}\rangle$ requires to consider the algebra and the
eigenvalues of the {\it global} Poincar\'e operators $P_\alpha$,
$M_{\alpha,\beta}$ ( $\alpha$ ,$\beta$=0, 1, 2, 3) where $P_\alpha$
are the 4 generators of the space-time translations and
$M_{\alpha\beta}=-M_{\beta\alpha}$ are the  6 generators of the
Lorentzian rotations with commutation relations
\begin{equation} \label{tras1} [P_\alpha,P_\beta]=0 \end{equation}
\begin{equation} \label{boost} [M_{\alpha\beta}, P_\gamma]=
\eta_{\beta\gamma}P_\alpha - \eta_{\alpha\gamma}P_\beta
\end{equation} \begin{equation} \label{tras2} [M_{\alpha\beta},
M_{\gamma\delta}]= \eta_{\alpha\gamma}M_{\beta\delta}+
\eta_{\beta\delta}M_{\alpha\gamma}
-\eta_{\beta\gamma}M_{\alpha\delta}-\eta_{\alpha\delta}M_{\beta\gamma}
\end{equation} where $\eta_{\alpha\beta}={\rm diag}(1,-1,-1,-1)$.
In this framework, as first discussed in refs.\cite{epjc,dedicated},
exact Lorentz invariance of the vacuum requires to impose the
problematic condition of a vanishing vacuum energy. As an example,
one can consider the generator of a Lorentz-transformation along the
1-axis ${M}_{01}$ for which one finds \BE
P_1{M}_{01}|\Psi^{(0)}\rangle={M}_{01}P_1|\Psi^{(0)}\rangle
+P_0|\Psi^{(0)}\rangle \EE Therefore, even assuming zero spatial
momentum for the vacuum condensation phenomenon, a non zero vacuum
energy $E_0$ implies \BE P_1{M}_{01}|\Psi^{(0)}\rangle=
E_0|\Psi^{(0)}\rangle \neq 0\EE This means that the state
${M}_{01}|\Psi^{(0)}\rangle$ is non vanishing so that the reference
vacuum state $|\Psi^{(0)}\rangle$ cannot be Lorentz invariant.

The simplest consequence of such non-invariance of the vacuum is an
energy-momentum flow along the direction of motion with respect to
$\Sigma$. In fact, by defining a boosted vacuum state $|
\Psi'\rangle$ as
\begin{equation}\label{trasf1} | \Psi'\rangle=
e^{\lambda'{M}_{01}}|\Psi^{(0)}\rangle \end{equation} (recall that
${M}_{01} \equiv -i{L}_1$ is an anti-hermitian operator) and using
the relations \begin{equation} \label{po1}
e^{-\lambda'{M}_{01}}~{P}_1~
e^{\lambda'{M}_{01}}=\cosh\lambda'~{P}_1 + \sinh\lambda'~{P}_0
\end{equation} \begin{equation} \label{po2} e^{-\lambda'{M}_{01}}~{P}_0
~e^{\lambda'{M}_{01}}=\sinh\lambda'~{P}_1 + \cosh\lambda'~{P}_0
\end{equation} one finds\begin{equation}
 \langle {{P}_1}\rangle_{\Psi'}=E_0\sinh\lambda' ~~~~~~~~~~~~~~~
 \langle {{P}_0}\rangle_{\Psi'}=E_0\cosh\lambda'\end{equation}
Clearly this result contrasts with the alternative approach where
one tends to consider $E_0$ as a spurious concept and rather tries
to characterize the vacuum through a {\it local} energy-momentum
tensor of the form \cite{zeldovich,weinberg}
\begin{equation}\label{zeld} \langle {W}_{\mu\nu}\rangle_
{\Psi^{(0)}}=\rho_v ~\eta_{\mu\nu}\end{equation} ($\rho_v$ being a
space-time independent constant). In this case, one is driven to
completely different conclusions. In fact, by introducing the
Lorentz transformation matrices $\Lambda^\mu_\nu$ to any moving
frame $S'$, defining $\langle {W}_{\mu\nu}\rangle_{\Psi'}$ through
the relation
\begin{equation}\langle \label{cov}
{W}_{\mu\nu}\rangle_{\Psi'}=\Lambda^{\sigma}{_\mu}\Lambda^{\rho}{_\nu}
~\langle{W}_{\sigma\rho}\rangle_{\Psi^{(0)}}\end{equation} and using
Eq.(\ref{zeld}), the expectation value of ${W}_{0i}$ in any boosted
vacuum state $| \Psi'\rangle$ vanishes, just as it vanishes in
$|\Psi^{(0)}\rangle$, so that \begin{equation} \label{density1} \int
d^3x~ \langle {W}_{0i}\rangle_{\Psi'} \equiv \langle
{{P}_i}\rangle_{\Psi'}= 0 \end{equation} Still, the idea to simply
get rid of $E_0$ gives rise to some problems. For instance, in a
second-quantized formalism, single-particle energies $E_1({\bf{p}})$
are defined as the energies of the corresponding one-particle states
$|{\bf{p}}\rangle$ minus the energy of the zero-particle, vacuum
state. If $E_0$ is considered a spurious concept,  $E_1({\bf{p}})$
will also become an ill-defined quantity. At the same time, the idea
to characterize the physical vacuum  through its energy $E_0$ has
solid motivations. The ground state, in fact, is by definition the
state with lowest energy as obtained from the solution of a minimum
problem. As such, it should correspond to an energy eigenstate in
view of the standard equivalence between eigenvalue equation and
Rayleigh-Ritz variational procedure.

Finally, at a deeper level, one should also realize that in an
approach based solely on Eq.(\ref{zeld}) the properties of
$|\Psi^{(0)}\rangle$ under a Lorentz transformation are not well
defined. In fact, a transformed vacuum state $| \Psi'\rangle$ is
obtained, for instance, by acting on $|\Psi^{(0)}\rangle$ with the
boost generator ${M}_{01}$. Once $|\Psi^{(0)}\rangle$ is considered
an eigenstate of the energy-momentum operator, one can definitely
show that, for $E_0\neq 0$, $| \Psi'\rangle$ and
$|\Psi^{(0)}\rangle$ differ non-trivially. On the other hand, if
$E_0=0$ there are only two alternatives: either
${M}_{01}|\Psi^{(0)}\rangle=0$, so that
$|\Psi'\rangle=|\Psi^{(0)}\rangle$,  or ${M}_{01}|\Psi^{(0)}\rangle$
is a state vector proportional to $|\Psi^{(0)}\rangle$, so that $|
\Psi'\rangle$ and $|\Psi^{(0)}\rangle$ differ by a phase factor.
Therefore, if the structure in Eq.(\ref{zeld}) were really
equivalent to impose the exact Lorentz invariance of the vacuum, it
should be possible to show similar results, for instance that such a
$|\Psi^{(0)}\rangle$ state can remain invariant under a boost, i.e.
be an eigenstate of \BE {M}_{0i}=-i\int d^3x~(x_i{W}_{00}-x_0
{W}_{0i}) \EE with zero eigenvalue. However, there is no way to
obtain such a result by just starting from Eq.(\ref{zeld}) (this
only amounts to the weaker condition $\langle {M}_{0i}\rangle_
{\Psi^{(0)}}=0$). Thus, it should not come as a surprise that one
can run into contradictory statements and it is not obvious that the
local relations (\ref{zeld}) represent a more fundamental approach
to the vacuum.

While a non-zero vacuum energy $E_0\neq 0$ might have different
explanations, one should also be aware that, in interacting quantum
field theories, there is no known way to ensure consistently the
condition $E_0=0$ without imposing an {\it unbroken supersymmetry}
(which is not phenomenologically acceptable). This makes the issue
of an exact Lorentz invariant vacuum a difficult problem which, at
present, cannot be solved on purely theoretical grounds
\footnote{One could also argue that a satisfactory solution of the
vacuum energy problem lies definitely beyond flat space.
Nevertheless, in the absence of a consistent quantum theory of
gravity, physical models of the vacuum in flat space can be useful
to clarify a crucial point that, so far, remains obscure: the huge
difference which is seen when comparing the typical vacuum-energy
scales of particle physics with the value of the cosmological term
needed in Einstein's equations to fit the observations. As discussed
in Sect.5, in this perspective `emergent-gravity' approaches
\cite{barcelo1,barcelo2,volo,bosegravity}, where gravity somehow
arises from long-wavelength excitations of the same physical
flat-space vacuum, may become natural and, to find the appropriate
infinitesimal value of the cosmological term, one is naturally lead
\cite{jannes,cosmo} to sharpen our understanding of the vacuum
structure and of its excitation mechanisms by starting from the
picture of a superfluid medium in flat space.}.

\vfill\eject

\end{document}